\documentclass[fleqn,usenatbib]{mnras}

\usepackage{newtxtext,newtxmath}

\usepackage[T1]{fontenc}
\usepackage{ae,aecompl}

\usepackage{graphicx}	% Including figure files
\usepackage{amsmath}	% Advanced maths commands
\usepackage{amssymb}	% Extra maths symbols
\usepackage[T1]{fontenc}      % Para PT-BR
\usepackage[utf8]{inputenc}   % Para PT-BR

\usepackage{mathtools}
\usepackage{verbatim}

\usepackage{hyphenat}         % Para PT-BR
\newcommand{\NIFS}{{\tt {NIFS}}}
\newcommand{\hb}{H$\beta$}

\newcommand{\kms}{~\textrm{km}\,\textrm{s}$^{-1}$}

\title[Gas inflows in the polar ring of NGC\,4111]{Gas inflows in the polar ring of NGC\,4111: the birth of an AGN}

\author[Hauschild-Roier et al.]{
Gabriel R. Hauschild-Roier,$^{1}$\thanks{E-mail: gabriel.roier@ufrgs.br}
Thaisa Storchi-Bergmann,$^{1}$
Richard M. McDermid,$^{2}$
\newauthor
Jonelle L. Walsh,$^{3}$
Joanne Tan,$^{3}$
Jonathan Cohn,$^{3}$
Davor Krajnović,$^{4}$
Jenny Greene,$^{5}$
\newauthor
Monica Valluri,$^{6}$
Kayhan Gültekin,$^{6}$
Sabine Thater,$^{7}$
Glenn van de Ven,$^{7}$
Karl Gebhardt,$^{8}$
\newauthor
Nora Lützgendorf,$^{9}$
Benjamin D. Boizelle,$^{3, 10}$
Chung-Pei Ma$^{11}$
and Aaron J. Barth$^{12}$
\\
\\
$^{1}$Instituto de Física, Universidade Federal do Rio Grande do Sul, Av. Bento Gonçalves 9500, 91501-970, Porto Alegre, RS, Brazil\\
$^{2}$Astronomy, Astrophysics, and Astrophotonics Research Centre, Department of Physics and Astronomy, Macquarie University, NSW 2109, Australia\\
$^{3}$George P. and Cynthia W. Mitchell Institute for Fundamental Physics and Astronomy, Texas A\&M University, 4242 TAMU, College Station, TX 77843, USA\\
$^{4}$Leibniz Institute for Astrophysics, Potsdam, Germany\\
$^{5}$Department of Astrophysical Sciences, Princeton University, NJ, USA\\
$^{6}$Department of Astronomy, University of Michigan, Ann Arbor, MI 48109, USA\\
$^{7}$Department of Astronomy, University of Vienna, Austria\\
$^{8}$Department of Astronomy, University of Texas at Austin, USA\\
$^{9}$European Space Agency, c/o STScI, 3700 San Martin Drive, 21218 Baltimore, MD, USA\\
$^{10}$Department of Physics and Astronomy, N283 ESC, Brigham Young University, Provo, UT 84602, USA\\
$^{11}$Department of Astronomy, UC Berkeley, USA\\
$^{12}$Department of Physics and Astronomy, 4129 Frederick Reines Hall, University of California, Irvine CA 92697-4575, USA\\
}

\date{Accepted XXX. Received YYY; in original form ZZZ}

\pubyear{2022}

\begin{document}
\label{firstpage}
\pagerange{\pageref{firstpage}--\pageref{lastpage}}
\maketitle

\begin{abstract}

We have used Hubble Space Telescope (HST) images, SAURON Integral Field Spectroscopy (IFS) and adaptative optics assisted Gemini NIFS near-infrared K-band IFS to map the stellar and gas distribution, excitation and kinematics of the inner few kpc of the nearby edge-on S0 galaxy NGC\,4111. The HST images map its $\approx$\,450\,pc diameter dusty polar ring, with an estimated gas mass $\ge10^7$\,M$_\odot$. The NIFS datacube maps the inner 110\,pc radius at $\approx$\,7\,pc spatial resolution revealing a $\approx$\,220\,pc diameter polar ring in hot (2267$\pm166$\,K) molecular H$_2$ 1-0 S(1) gas embedded in the polar ring. The stellar velocity field shows disk-dominated kinematics along the galaxy plane both in the SAURON large-scale and in the NIFS nuclear-scale data. The large-scale [\ion{O}{iii}] $\lambda5007$\,\AA\ velocity field shows a superposition of two disk kinematics: one similar to that of the stars and another along the polar ring, showing non-circular motions that seem to connect with the velocity field of the nuclear H$_2$ ring, whose kinematics indicate accelerated inflow to the nucleus. The estimated mass inflow rate is enough not only to feed an Active Galactic Nucleus (AGN) but also to trigger circumnuclear star formation in the near future. We propose a scenario in which gas from the polar ring, which probably originated from the capture of a dwarf galaxy, is moving inwards and triggering an AGN, as supported by the local X-ray emission, which seems to be the source of the H$_2$ 1-0 S(1) excitation. The fact that we see neither near-UV nor Br$\gamma$ emission suggests that the nascent AGN is still deeply buried under the optically thick dust of the polar ring.

\end{abstract}


%%%%%%%%%%%%%%%%%%%%%%%%%%%%%%%%%%%%%%%%%%%%%%%%%%

%%%%%%%%%%%%%%%%% BODY OF PAPER %%%%%%%%%%%%%%%%%%
\section{Introduction}

Interactions between galaxies impact their evolution and that of their central supermassive black holes (SMBHs), triggering episodes of star formation and nuclear activity.
%are events that impact their evolution and of  their nuclear Supermassive Black Holes (SMBHs), being triggers of episodes of star formation and nuclear activity in galaxies. 
Major mergers seem to be the dominant process of SMBH growth at large SMBH masses ($\ge\,10^8$ M$_{\odot}$), triggering luminous Active Galactic Nuclei (AGN), while at lower AGN luminosities ($\le 10^{44}$\,erg\,s$^{-1}$), the most probable triggers are minor mergers and secular processes driving inflows towards the nucleus such as gravitational torques in nuclear spirals and bars \citep[][and references therein]{storchi-bergmann_observational_2019}.
Minor mergers are frequent in dense galactic environments, when massive early-type galaxies capture gas-rich dwarf galaxies that replenishes the inner region of the galaxy with gas, leading to episodes of renewed star formation and/or the triggering of nuclear activity \citep[e.g.][]{neistein_what_2014}. 

Such a minor merger may have recently happened in the nearby SO edge-on galaxy NGC\,4111, known to host an extended H\,I envelope \citep{wolfinger_blind_2013}. A member of the Ursa Major galaxy group, this galaxy lies at a distance of 15.1 Mpc (according to the NASA Extragalactic Database (NED)\footnote{https://ned.ipac.caltech.edu/}, using a $\Lambda\text{CDM}$ cosmology with $H_0 = 67.8$ km s$^{-1}$ Mpc$^{-1}$, $\Omega_m = 0.308$ and $\Omega_\Lambda = 0.692$) -- for which 1$^{\prime\prime}$ corresponds to 73.1\,pc. NGC\,4111 has at least 4 companion galaxies located within 250\,kpc from it, with the nearest only 30-40\,kpc away \citep{karachentsev_anatomy_2013, pak_properties_2014, kasparova_diversity_2016}. NGC\,4111 has a dust structure crossing the nucleus perpendicular to the galaxy disk \citep{barth_search_1998} associated with H\,I filaments, which suggest an origin from tidal stripping of gas from a nearby or recently captured galaxy \citep{verheijen_hi_2001, verheijen_galaxy_2004, pak_properties_2014}.

\citet{kasparova_diversity_2016} showed that NGC\,4111 has a flat stellar age distribution of $\sim$\,5\,Gyr across most of its disk. In the inner region, however, with galaxy midplane distance of $\sim$\,700\,pc, single stellar population modelling reveal a more metal-rich and younger stellar population with a characteristic 4\,Gyr age. This very flattened young stellar component could be due to a previous gas-rich minor merger.

NGC\,4111 was identified as an AGN by \citet{gonzalez-martin_x-ray_2009} using Chandra X-ray data. In that work, the column density of HI and H$_2$ were calculated to be $4.67 \times 10^{22}$ cm$^{-2}$ and $37.71 \times 10^{22}$ cm$^{-2}$, respectively. NGC\,4111 images from Chandra show extended soft X-ray emission and nuclear-only hard emission, with luminosities measured as $\log(L_\mathrm{soft})$ = 40.9 erg s$^{-1}$ (0.5 - 2.0 keV) and $\log(L_\mathrm{hard})$ = 40.4 erg s$^{-1}$ (2.0 - 10.0 keV). This suggests that the source of high energy X-rays is a fairly obscured AGN which is embedded in the dense dust ring, while the soft emission is spread across the galaxy plane and has a maximum in the central region. This is also supported by the fact that near-UV emission from the nucleus is not detected in HST images \citep{barth_search_1998}. NGC\,4111 shows 5\,GHz radio emission extended by 110.4\,pc$\,\times$\,72.2\,pc and position angle of 29$\pm$11$^\circ$ according to \citet{nyland_atlas3d_2016}; however, the origin of the radio emission is unclear and the nucleus of the galaxy is classified as a LINER \citep{ho_search_1997}.
 
 NGC\,4111 was observed with the Gemini Near-Infrared Spectrograph (NIFS) as part of the project ``Addressing a Bias in the Relation Between Galaxies and Their Central Black Holes" (P.I. Jonelle Walsh). While the main goal of the project is to determine SMBH masses from the stellar kinematics in nearby galaxies, the NIFS K-band observations of NGC\,4111 revealed many lines of hot molecular H$_2$ gas emission in the spectra. These lines reveal peculiar excitation and kinematics that deserve a study of its own. The goal of the present paper is to investigate the nature of the excitation and kinematics of the hot H$_2$ gas emission of the inner $\approx$\,100\,pc of the galaxy. In order to do this, we also employed Hubble Space Telescope (HST) images and large scale integral field data from the Spectrographic Areal Unit for Research on Optical Nebulae survey (SAURON).
 
 This paper is organised as follows. In Section 2 we present the observations and data reduction. In Section 3, we present the measurements for the HST (regarding the photometry), SAURON and NIFS data (regarding the emission lines fitting and kinematics, and the stellar kinematics). In Section 4, we present our results and discussion for all 3 instruments, so that in Section 5 we compare these results and propose a scenario for the observational evidences of NGC\,4111. Finally, in Section 7 we present our conclusions.

\section{Observations}

As a part of GO-15323 (PI: Jonelle Walsh), we obtained HST Wide Field Camera 3 (WFC3) imaging in the F475W (see Figure\,\ref{fig:1}) and F160W filters. The F160W data were obtained using the RAPID sequence in a 2-point WFC3-IR-DITHER-LINE pattern, with a total 53\,s exposure time. The F475W data were obtained in an ideal 4-point dither pattern using only the UVIS2-2K2C-SUB aperture, with a total 920\,s exposure time. The WFC3 data were combined into separate F475W and F160W mosaics with the same origin and pixel scale using {\tt AstroDrizzle} \citep{gonzaga_drizzlepac_2012}.

K-band spectroscopy of the nuclear region of NGC\,4111 was obtained using the instrument NIFS -- Near-Infrared Integral Field Spectrograph \citep{mcgregor_gemini_2003}, with the adaptive optics module Altair \citep{herriot_progress_2000, boccas_laser_2006} at the Gemini North Telescope on 2019 May 9. We acquired 600\,s exposures following an object-sky-object sequence, with a total on-source time of 1 hour via the program GN-2019A-LP-8 (P.I. Jonelle Walsh). We used the $H+K$ filter and $K$ grating centered on $2.20\,\mu$m, which provided coverage of 1.99 $\,\mu$m - 2.40 $\,\mu$m with $R \sim 5290$. NIFS is an image slicer with a Field-of-View (FoV) of 3$^{\prime\prime}\times3^{\prime\prime}$ -- corresponding to $219\times219$\,pc$^2$ at the galaxy, sampled with pixels of $0\farcs04\times0\farcs1$; the observations with the adaptative optics module Altair delivered data with a Point-Spread Function (PSF) -- as determined from the spatial profiles of standard stars --  FWHM of $0\farcs1$ corresponding to a spatial resolution of $\approx$\,7\,pc at NGC\,4111. We reduced the NIFS data using IRAF tasks \citep{tody_iraf_1986, tody_iraf_1993} within the Gemini package v1.14, in combination with Python scripts we developed \footnote{https://github.com/jlwalsh12/NIFS-reduction-pipeline} based on the example NIFS processing scripts \footnote{https://www.gemini.edu/instrumentation/nifs/data-reduction}. The main steps included sky subtraction, flat-fielding, cosmic ray cleaning and bad pixel removal, wavelength calibration, and spatial rectification. The galaxy spectra were telluric corrected using an A0 V star (HD95126 or HD116405), after removal of the Br$\gamma$ absorption line and the blackbody continuum. We constructed a temporary data cube for each galaxy exposure and collapsed the cubes to determine the spatial offsets by means of a two-dimensional cross-correlation. Then all six of the galaxy frames were interpolated onto a single data cube with 0\farcs05 vs. 0\farcs05 spaxels and wavelength sampling of 2.13\AA. Finally, a rough flux calibration was carried out using the telluric-corrected HD116405 spectrum and comparing the flux density at the isophotal wavelength of the Two Micron All Sky Survey (2MASS) $Ks$ filter to the 2MASS magnitude reported in the SIMBAD astronomical database \citep{wenger_simbad_2000}. This conversion was applied to the final NGC\,4111 data cube, resulting in calibrated units of erg s$^{-1}$ cm$^{-2}$ \AA$^{-1}$. We reduced the sky frames in a similar manner, but subtracted a 600\,s master dark frame. We applied the same spatial offsets as the ones used for the galaxy frames to produce a single sky cube and measured the line spread function over the NIFS field-of-view from 13 strong, isolated sky lines. The sky lines spectral resolution varies over the NIFS field-of-view, ranging from a FWHM of 3.5 to 5.9\,\AA, with a FWHM median resolution of 3.9\,\AA.

Optical integral field spectroscopy (IFS) data were obtained with the SAURON instrument \citep{bacon_sauron_2001} on the William Herschel Telescope, La Palma as part of the ATLAS$^{\textrm{ 3D}}$ Survey \citep{cappellari_atlas3d_2011}. SAURON provides a 31\arcsec$\times$41\arcsec\/ field of view -- corresponding to 2.3\,kpc$\times$3.0\,kpc at the galaxy, sampled with 0.94\arcsec\/ square lenslets, and with a spectral resolution of 4.2\,\AA\/ FWHM ($\sigma_{\mathrm{inst}} = 105$\kms), covering the wavelength range 4800–5380\,\AA. The observations comprised the combination of two exposures of 30\,minutes, obtained on UT 2007 April 20, together with associated calibrations. The data were reduced via standard procedures using the dedicated {\tt XSAURON} software package \citep{bacon_sauron_2001, emsellem_sauron_2004}. The approximate seeing FWHM was 1.2\arcsec, corresponding to a spatial resolution of $\approx$\,88\,pc at the galaxy. These reduced spectral data cubes are publicly available at www.purl.org/atlas3d.

The aforementioned data were astrometrically matched using the WCS data in the headers, followed by a visual inspection to assure the correct relative position of the images.

\section{Measurements}

\subsection{HST}

\begin{figure}
	\includegraphics[width=\columnwidth]{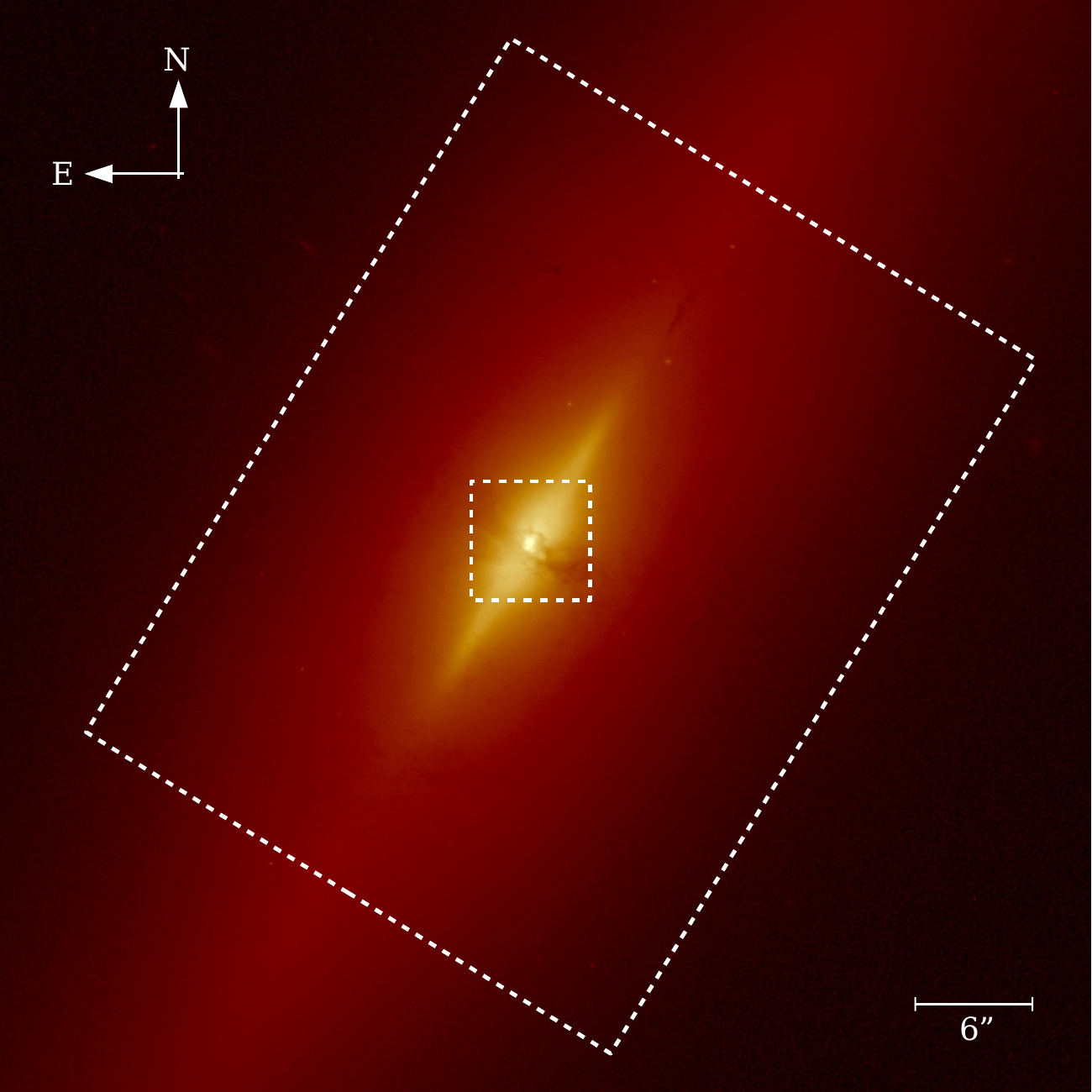}
    \caption{HST large scale image of NGC\,4111 in the F475W filter covering a 4.8\,kpc$\times$4.8\,kpc region. The white dashed-line rectangle shows the $31^{\prime\prime}\times41^{\prime\prime}$ SAURON FoV and the smaller white dashed-line square delimits a $6^{\prime\prime}\times6^{\prime\prime}$ region containing the dusty polar ring shown in the colour map of Figure \ref{fig:colourmap}. North is up and East is left.}
    \label{fig:1}
\end{figure}

\begin{figure}
    \centering
    \includegraphics[width=\columnwidth]{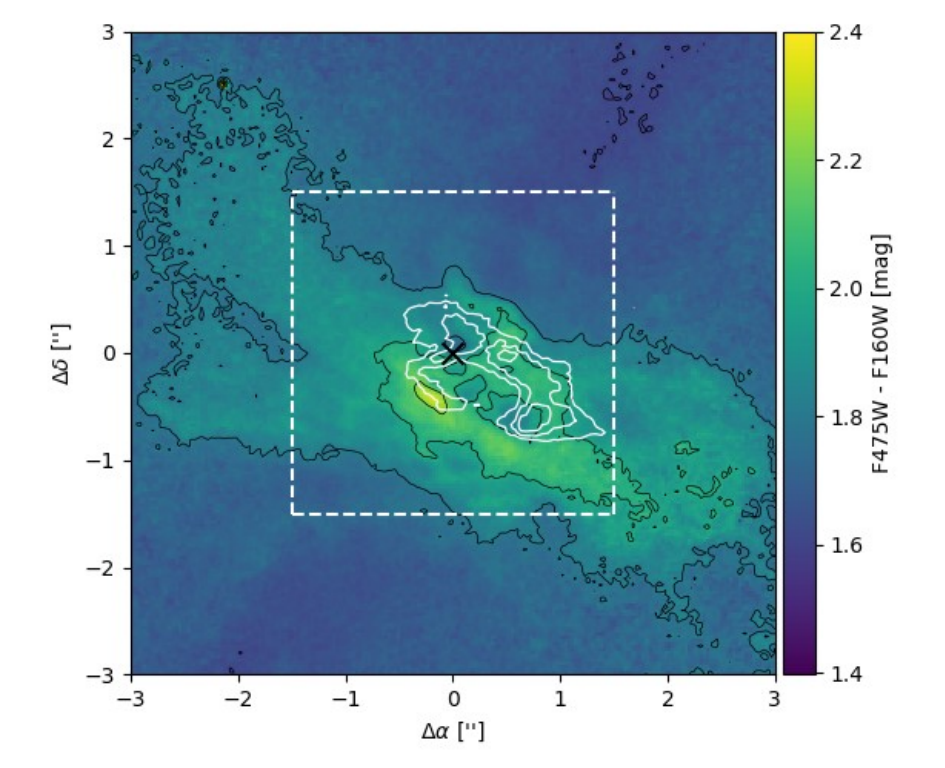}
    \caption{HST F475W - F160W colour map within the inner 6$^{\prime\prime}\times6^{\prime\prime}$ (440$\times$440\,pc$^2$ at NGC\,4111). Overplotted in white are the contours from the H$_2 \lambda2.1218\,\mu$m emission line flux distribution (see Sec.\,\ref{Sec:NIFS_spectra}), and in black a cross shows the location of the peak of the stellar K-band continuum from NIFS, adopted as corresponding to the galaxy nucleus. The white dashed lines demarcate the NIFS FoV.}
    \label{fig:colourmap}
\end{figure}

Figure\,\ref{fig:1} shows the HST F475W continuum image with the 31$^{\prime\prime}\times41^{\prime\prime}$ SAURON FoV delineated by a white dashed-line rectangle. This image shows the edge-on orientation of the galaxy and the dusty polar ring crossing the nucleus approximately perpendicular to the disk of the galaxy. The 6$^{\prime\prime}\times6^{\prime\prime}$ white dashed-line square encloses the region where the dusty polar ring is observed. 

Using the F475W and the F160W images, we built the F475W\,-\,F160W colour map, with the inner 6$^{\prime\prime}\times6^{\prime\prime}$ region shown in Fig.\,\ref{fig:colourmap}. This map shows that the dusty ring has a diameter of $\approx$\,450\,pc ($6\farcs3$) and a width of $\approx$\,130\,pc ($1\farcs8$). The 3$^{\prime\prime}\times3^{\prime\prime}$ dashed white square in this figure shows the NIFS FoV, while the \textbf{white} contours delineates the flux distribution in the H$_2$(1-0)\,S(1)\,$\lambda$\,2.1218$\,\mu$m emission line that has the shape of an off-centered inclined ring, as shown and discussed in the following section (see Fig. \ref{fig:h2ring}).

\subsection{SAURON}
\label{Sec:SAURON_spectra}

The SAURON optical spectra were analysed first by using Voronoi binning \citep{cappellari_adaptive_2003} to increase the SNR of the spectra. We adopted a lower threshold of SNR=40 and ran the pPXF code \citep{cappellari_parametric_2004} on the binned spaxels to measure the stellar kinematics. pPXF uses a stellar spectral base to fit the absorption lines in the galaxy spectrum in order to measure the kinematics (velocity, velocity dispersion and the $h_3$ and $h_4$ Gauss-Hermite polynomials coefficients) in these lines. In this work, we used the MILES stellar templates \citep{falcon-barroso_updated_2011} as a base for pPXF. Emission line properties were extracted using an implementation of the GANDALF (Gas AND Absorption Line Fitting) software \citep{sarzi_gandalf_2017}. In this implementation, the previously-derived pPXF stellar kinematics are fixed, and MILES SSP model templates are broadened by the corresponding LOSVD. The template weights are permitted to vary, together with the parameters of single Gaussian emission line components. The [\ion{O}{iii}]$\lambda\lambda 4959,5007$ doublet is initially fitted, as this is generally the most easily detected line without strong coincident absorption features. This fixes the emission line kinematics for a second pass, where an \hb\/ emission component is also included. The fitted emission line is considered a secure detection if it has an amplitude 4 times higher than the local noise level, otherwise it is ignored.

\subsection{NIFS}
\label{Sec:NIFS_spectra}

\begin{figure*}
    \centering
    \includegraphics[width=\textwidth]{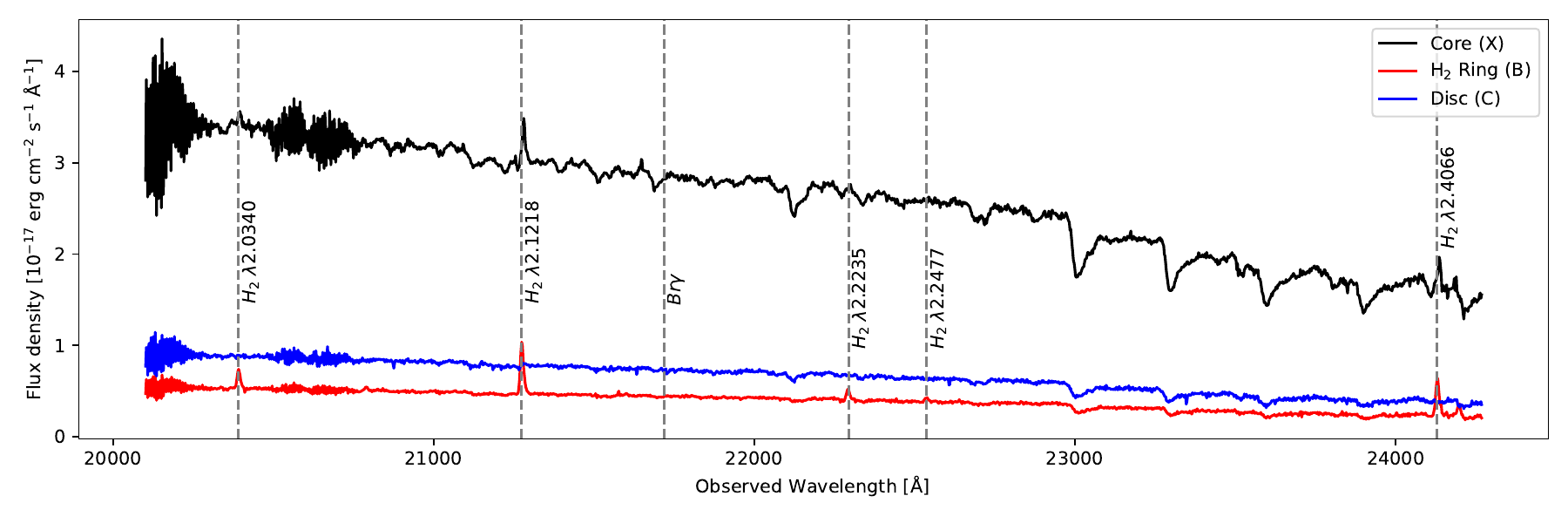}
    \caption{NIFS spectra in observed wavelength from 3 representative regions of the galaxy, integrated within apertures of 0\farcs1$\times$0\farcs1: nuclear spectrum in black; H$_2$ ring region spectrum in red, extracted from position \textit{B} in Figure \ref{fig:h2ring}; spectrum from the disk of the galaxy in blue, extracted from position \textit{C} in Figure \ref{fig:h2ring}. The wavelengths of the H$_2$ emission lines are marked with dashed grey lines, along with that of the absent Br$\gamma$ line.}
    \label{fig:spec}
\end{figure*}

\begin{figure}
    \centering
    \includegraphics[width=\columnwidth]{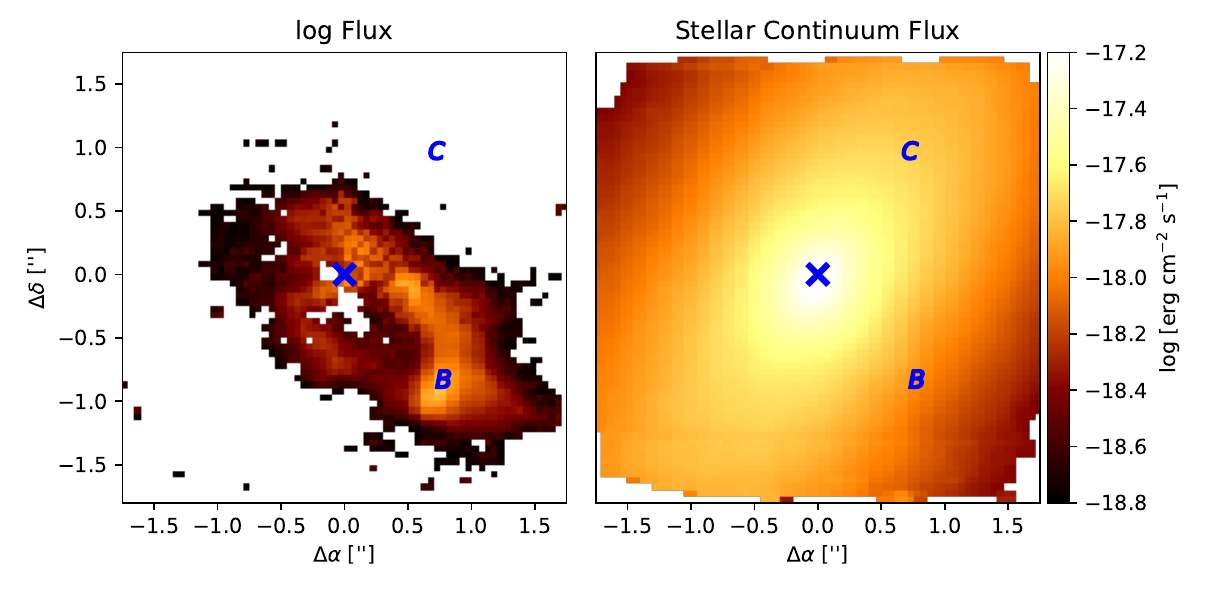}
    \caption{\textbf{Left:} Integrated flux distribution in the molecular H$_2$(1-0) S(1) $\lambda2.1218\,\mu$m emission line. \textbf{Right:} Stellar continuum flux distribution from the NIFS datacube, obtained by integrating the spectra within the 2.175-2.200\,$\mu$m wavelength range. The cross shows the spaxel with the highest integrated continuum flux -- identified with the galaxy nucleus; the letters \textit{B} and \textit{C} identify the locations from where the spectra identified in Figure \ref{fig:spec} were extracted.}
    \label{fig:h2ring}
\end{figure}

Sample spectra from NIFS are shown in Figure \ref{fig:spec}, integrated within 0\farcs1$\times$0\farcs1 windows, corresponding to the nucleus (labeled \textit{X} in Fig.\,\ref{fig:h2ring}) and two extranuclear regions: one from a position over the H$_2$ off-centered ring (labelled \textit{B} in Fig. \ref{fig:h2ring}) and from a region beyond this ring at 1\farcs25 (91\,pc) from the nucleus (labelled \textit{C} in Fig.\,\ref{fig:h2ring}), comprising mostly light from the stellar component of the galaxy. The spectra show absorption lines from the stellar population, and H$_2$ emission lines from hot molecular gas. Fig.\,\ref{fig:spec} identifies the wavelengths of the H$_2$ emission lines as well as that of the Br$\gamma$ line, which is not present in the spectra. We note that the H$_2$ emission is observed only along the off-centered ring above and that the Br$\gamma$ line is absent over the whole area covered by the NIFS FoV (see Section \ref{sec:H2}).

\begin{figure}
    \centering
    \includegraphics[width=\columnwidth]{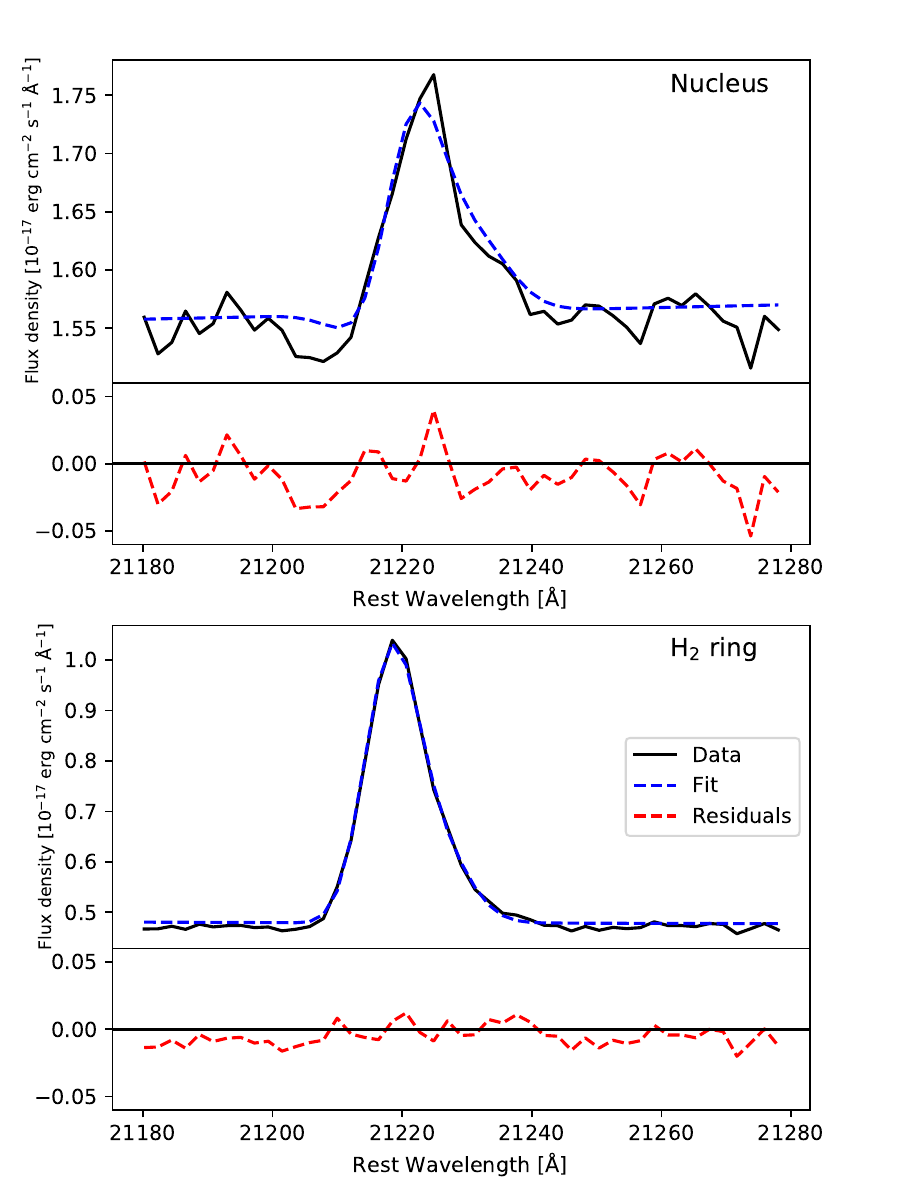}
    \caption{The fit of the H$_2\lambda2.1218\,\mu$m emission-line profile at the nucleus and position B in the ring (see Fig.\,\ref{fig:h2ring}).
    %regions shown in Figure\,\ref{fig:spec} in black, within the observed 21180\,-\,21280\,\AA\ spectral window used for fitting the molecular emission line with IFSCube. 
    The rest profile is shown in black lines and the IFSCube Gauss-Hermite polynomial fit in blue dashed lines, while the residuals are shown in red dashed lines.}
    \label{fig:h2_fit}
\end{figure}

In order to measure the properties of the emitting H$_2$ gas in the NIFS datacube, we used the latest version (v1.1) of the software IFSCube \citep{ruschel-dutra_danielrd6ifscube_2020} to fit the emission lines profiles with Gauss-Hermite polynomials, within the 21180\,-\,21280\,\AA\ spectral window (in observed wavelengths), which encompasses the H$_2$\,$\lambda$2.1218$\,\mu$m emission line and a continuum region. IFSCube fits a polynomial component to the continuum adjacent to the line, whose degree can be chosen during the fit. We opted by the use of a linear function, that fitted well the continuum in all spaxels. In Figure\,\ref{fig:h2_fit}, we present example fits from IFSCube for two spectra extracted in windows of $0\farcs1\times0\farcs1$, one from the nucleus and the other from a position along the molecular H$_2$ ring (position B in Fig.\ref{fig:h2ring}), showing in black lines the spectra, in dashed blue lines the fit and in dashed red lines the residuals. We imposed a lower threshold for the signal-to-noise ratio (SNR) as obtained from the ratio between the peak of the line and the noise in the continuum adjacent to the line (standard deviation of the continuum). Spaxels with a SNR less than 2 are regarded as noise and excluded from the fit. This same noise was used to apply the Monte Carlo technique in order to evaluate the errors in the parameters resulting from the fit. The left panel of Figure \ref{fig:h2ring} shows the resulting flux map of the H$_2$\,$\lambda$2.1218$\,\mu$m emission line, resembling a tilted ring oriented approximately perpendicular to the galaxy plane. The right panel shows the flux distribution of the stellar continuum, integrated in a window covering the 2.175-2.200$\,\mu$m wavelength range. In Fig.\,\ref{fig:h2ring} we also identify the nucleus and the other locations where the spectra from Figure \ref{fig:spec} were extracted.

\begin{figure}
    \centering
    \includegraphics[width=\columnwidth]{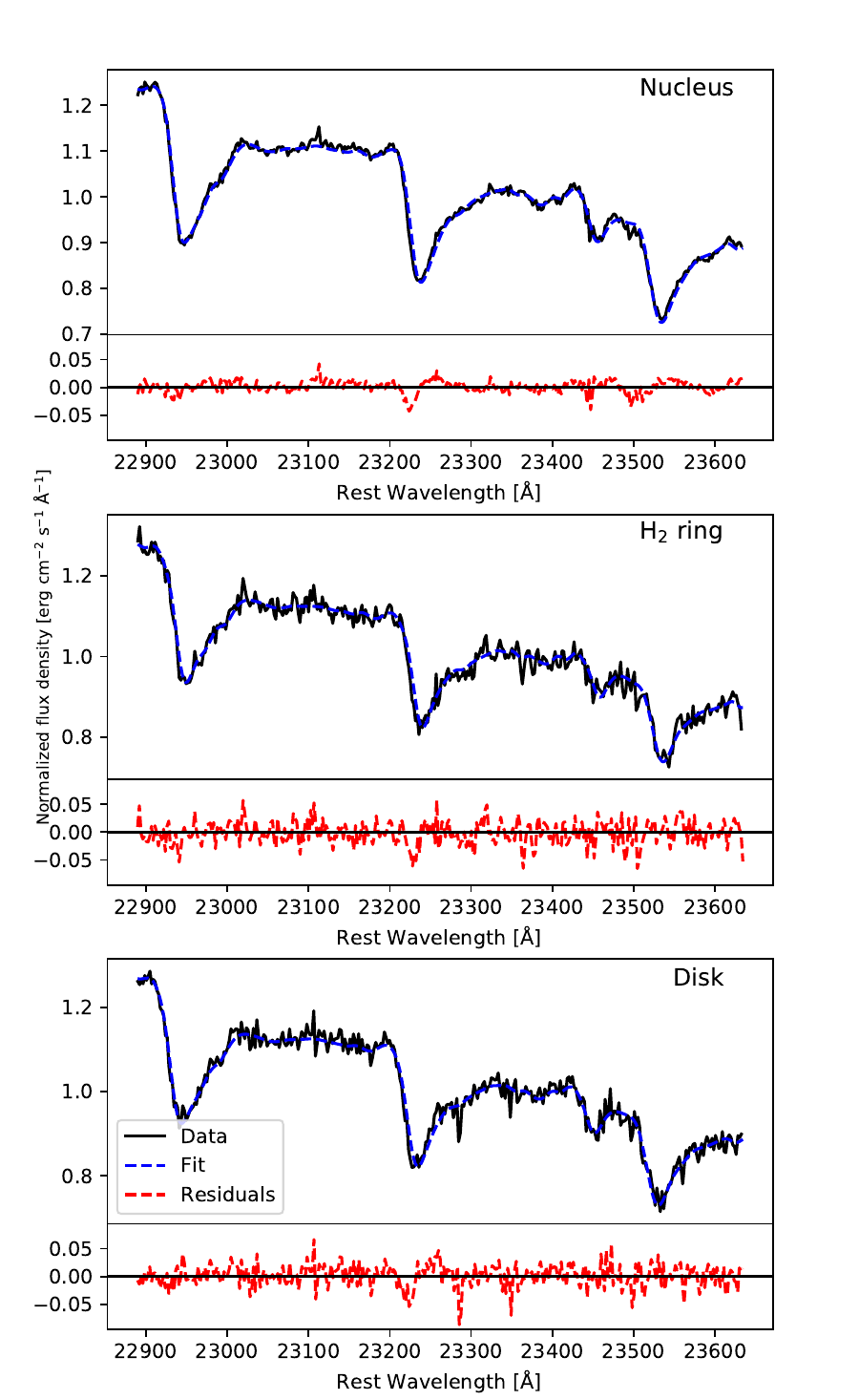}
    \caption{The fit of the stellar kinematics with pPXF was performed in the spectral range 22950-23700\AA (in observed wavelenghts). The figure shows three representative Voronoi-binned spectra in black, from the nucleus and positions B and C in Fig.\,\ref{fig:h2ring}, together with the fits in blue and residuals in red.}
    \label{fig:stellar_fit}
\end{figure}

We also obtained the stellar kinematics using a similar approach to that used for the SAURON data (see Section\,\ref{Sec:SAURON_spectra}): first we used Voronoi binning to increase the SNR of the spectra until it reached a lower threshold of SNR=50 in the stellar continuum and then executed pPXF with the Winge stellar templates \citep{winge_gemini_2009} as a base. The stellar kinematics fits at positions \textit{X}, \textit{B} and \textit{C} of Fig.\,\ref{fig:h2ring}, are shown, from top to bottom, in Figure\,\ref{fig:stellar_fit}, where the normalized spectra are shown in black, the fit in blue and the residuals in red.

\section{Results and Discussion}

\subsection{Galactic scale - HST}
\label{ss:hst_images}

The F475W - F160W colour map of Fig.\,\ref{fig:colourmap} traces the obscuring structure crossing the nuclear region seen in Fig.\,\ref{fig:1}, mapping its reddening and revealing the polar ring dust distribution which is extended perpendicular to the galaxy plane by approximately $\approx$\,450\,pc and shows the reddest colours in an off-centered ring with a major axis of $\approx$200\,pc. The NIFS FoV is identified via the dashed white square. We have overplotted on Fig.\,\ref{fig:colourmap}, contours from the the NIFS H$_2\ \lambda2.1218\,\mu$m flux map of Fig.\,\ref{fig:h2ring} shown in white,
while the location of the peak flux of the stellar continuum (adopted as the galaxy nucleus) is shonw as a black cross. This figure shows that the H$_2$ ring is spatially associated with the off-centered ring with the reddest colours in the polar ring that can be attributed to increased reddening by dust. This colour map reveals that the molecular hydrogen emission is \textit{de facto} related to the dust, with both presenting similar off-centered ring-like structure.

\subsubsection{Cold gas mass}

We have used the colour map of Fig.\,\ref{fig:colourmap} to obtain the extinction of the region under the assumption that the intrinsic colour of the stellar population is that of the galaxy disk, adopted as F475W - F160W$=1.5$, according to Fig.\,\ref{fig:colourmap}, and considering the dust obscuration as due to a foreground obscuring screen. Figure \ref{fig:av_map} of the Appendix shows the resulting A$_{F475W}-$A$_{F160W}$ extinction map of the inner 6\farcs0$\times$6\farcs0, that has an excess extinction at $\approx$4750\,\AA\ relative to $\approx$1.6$\,\mu$m of up to $\sim$ 0.8 mag in the ring relative to the stellar population. This extinction is consistent with the relatively high column density measured by \citet{gonzalez-martin_x-ray_2009} in the Chandra observations.

In our calculation, we have assumed that the dust is a foreground screen obscuring the light from the galaxy, and we have used the extinction law from \citet{savage_observed_1979} in order to derive the extinction values $E(B-V)$ from the A$_{F475W}-$A$_{F160W}$ extinction map, with A$_\lambda/E(B-V)$ values obtained from the last column of Table 2 of \citet{savage_observed_1979}. Also following \citet{savage_observed_1979}, we used the $E(B-V)$ values derived from the extinction map of Fig.\,\ref{fig:av_map} and a gas-to-colour excess of $\langle N(\ion{H}{i} +\mathrm{H}_2)/E(B-V)\rangle = 5.8 \times 10^{21}$ atoms cm$^{-2}$ mag$^{-1}$ in order to obtain a lower limit for the gas mass in each spaxel. Multiplying by the mass of the proton and dividing by the area of each spaxel in units of pc$^2$, we obtain the surface gas mass distribution shown in Figure \ref{fig:mass_hst} of the Appendix. Adding the mass values of all spaxels, the resulting lower limit for the cold gas mass in the 450\,pc polar ring is M$_{\text{gas}} = 9.8 \times 10^{6}$ M$_\odot$.

\subsection{Galactic scale - SAURON}

\subsubsection{Stellar and ionised gas flux distributions}

\begin{figure*}
    \centering
    \includegraphics[width=\textwidth]{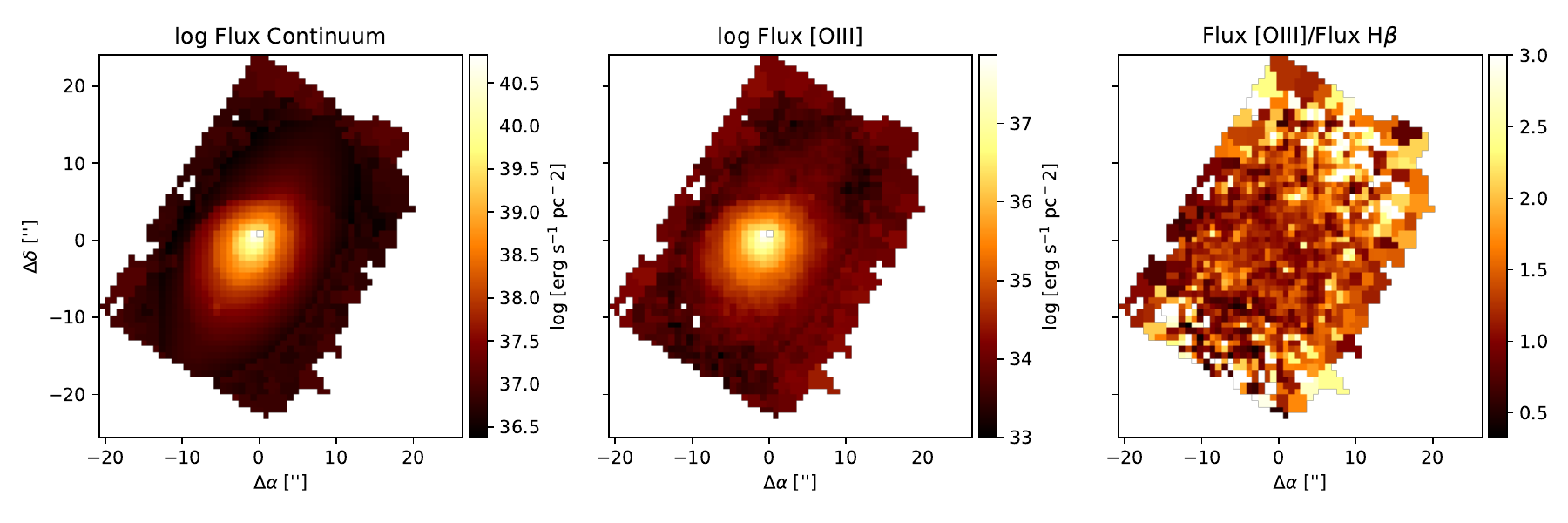}
    \caption{From left to right: Flux surface density distributions in the optical continuum, [\ion{O}{iii}]$\lambda$5007, and the line ratio map [\ion{O}{iii}]/H$\beta$ obtained from the SAURON data.}
    \label{fig:sauron_flux}
\end{figure*}

Flux distributions in the optical continuum, [\ion{O}{iii}]$\lambda$5007 and H$\beta$ gas emission lines over the inner $\approx$20$^{\prime\prime}$ of the galaxy obtained from the SAURON observations are shown in Fig.\,\ref{fig:sauron_flux}. In the continuum (mapping the optical stellar component), the flux distribution is brightest at the nucleus decreasing almost uniformly in all directions except for a slightly more abrupt decrease to the NW than to the SE. At faint levels there is some extension along the galaxy plane. The flux distributions in the emission lines are similar to that in the continuum for the highest flux levels around the nucleus but show also more emission outwards, in particular to the SW towards the border of the FoV.

The rightmost panel of Figure \ref{fig:sauron_flux} shows that the line flux maps of the [\ion{O}{iii}]$\lambda$5007 and H$\beta$ lines from the SAURON\ data produce a line ratio map with values in the range $1\le F$[\ion{O}{iii}]$/F(\mathrm{H\beta}) \le 3$, increasing towards the borders of the FoV. As the SAURON data does not cover the H$\alpha$ spectral region, we use a spectrum from this region from \citet{ho_search_1995} who find [F([\ion{N}{ii}]$\lambda6584$)/F(H$\alpha$)$\approx$1, which puts NGC\,4111 in the LINER region of the BPT diagram \citep{baldwin_classification_1981, kewley_theoretical_2001, kauffmann_host_2003, cid_fernandes_alternative_2010}, in agreement with the classification from \citet{nyland_atlas3d_2016}.

\subsubsection{Stellar kinematics}
\label{Sec:sauron_kin}

\begin{figure}
    \centering
    \includegraphics[width=\columnwidth]{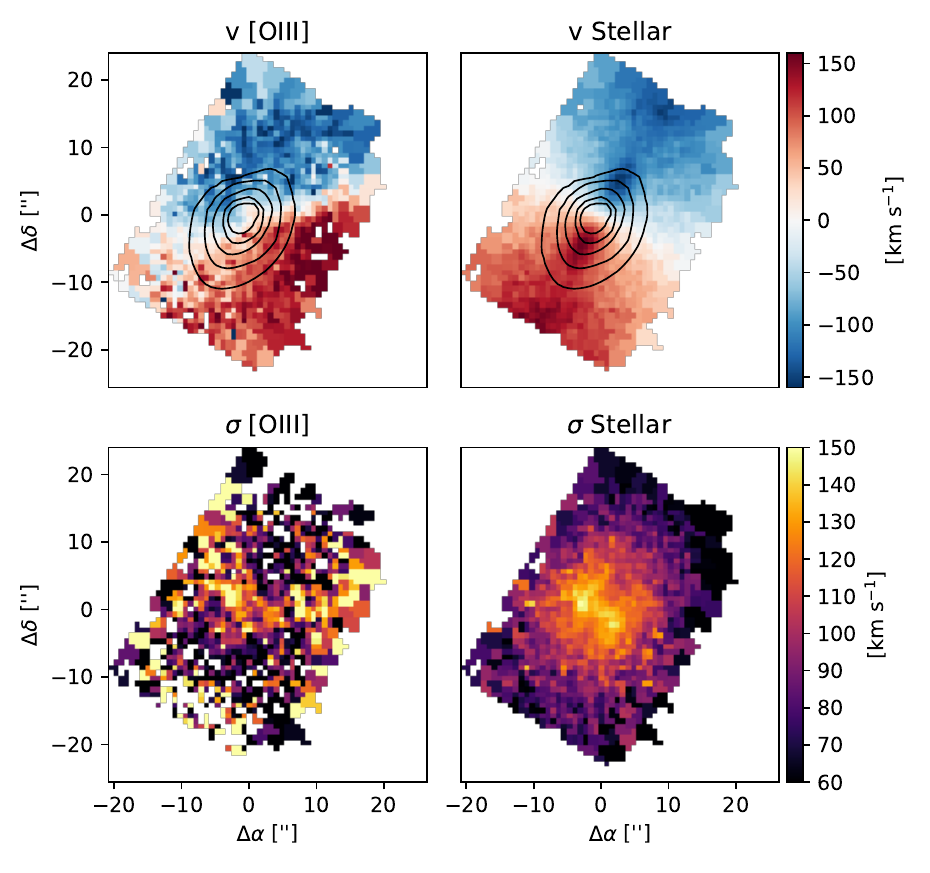}
    \caption{Kinematics from SAURON data. \textbf{Top:} ionised gas and stellar velocity field (centroid velocity $v$), with contour lines from the SAURON continuum flux from Fig. \ref{fig:sauron_flux}; \textbf{Bottom:} gas and stellar velocity dispersion $\sigma$.}
    \label{fig:sauron_kin}
\end{figure}

The SAURON data kinematics is shown in Fig. \ref{fig:sauron_kin}. The right panels show the stellar kinematics on kpc scales. The top panel shows a rotation pattern with redshifts to the SE and blueshifts to the NW. 

The large-scale stellar velocity field obtained from the SAURON data was previously analysed by \citet{krajnovic_atlas3d_2011}, who concluded that it presents ordered and disk-like rotation, with a kinematic position angle of $149.5\deg\pm2.2$. The stellar velocity dispersion $\sigma_\star$ %map from the SAURON\ data in Fig.\,\ref{fig:sauron_kin} 
shown in the bottom right panel, presents a small decrease in a linear structure, extended by $\approx$\,800\,pc crossing the galaxy bulge along the major axis of the galaxy. This feature suggests the presence of a colder structure -- as compared to the galaxy bulge, such as a disk of stars rotating in the galaxy plane that may have formed from recently acquired gas. Indeed, stellar population properties measured via line-strength indices and spectral fitting of this data by \citet{mcdermid_atlas3d_2015} reveal the presence of younger stars than that of the surrounding bulge in this region, what is also consistent with the stellar age of the region obtained by \citet{kasparova_diversity_2016}. 

%We discuss again the SAURON stellar velocity field again below (Sec.\,\ref{ss:kinemetry} using kinemetry \citep{krajnovic_kinemetry_2006} to characterize the stellar velocity field.

\subsubsection{Ionised gas kinematics}

The [\ion{O}{iii}] ionised gas kinematics probed by SAURON on kpc scales is shown in the left panels of Fig.\,\ref{fig:sauron_kin}. The ionised gas velocity field $v$[\ion{O}{iii}] shown in the top panel presents a different orientation from that of the stellar component, with redshifts to the S-SW and blueshifts to the N-NE. As compared with the stellar velocity field, it presents a $\approx$\,50$^\circ$ shift in the orientation of the apparent kinematic major axis from the $\approx150^\circ$ (PA east of North of the redshifted side) of the stellar kinematics, bringing the PA of the gas velocity field to $\approx$\,200$^\circ$. 

Furthermore, the ionised gas velocity dispersion map shows the highest values at various non-central locations. In particular, they seem to be observed mostly along the minor-axis, although not exactly and with a bend towards N on both the E and W sides of the galaxy. 

\subsubsection{Kinemetric analysis}
\label{ss:kinemetry}

In order to try to quantify the kinematic properties described above, we analyse the SAURON stellar and gaseous velocity maps using kinemetry \citep{krajnovic_kinemetry_2006}. This method searches for the best-fitting ellipses  (specified by the position angle PA and the ellipse flattening Q) along which the velocities can be described as a function of a cosine change in the eccentric anomaly. Kinemetry is built on the assumption that the velocities along the ellipses can be parameterised by $V = V_{\rm rot} \cos(\theta)$, where $V_{\rm rot}$ is the amplitude of rotation and $\theta$ is the eccentric anomaly. 

Kinemetry performs an harmonic decomposition of velocities along ellipses and determines the properties of the best-fitting ellipse by minimising the Fourier coefficients in a truncated Fourier Series, summing over the odd values of n: $V(r,\theta) = V_0 + \sum_{n=1}^3[ a_n(r) \sin(n\theta) + b_n(r) \cos(n\theta)]$, except for $b_1$ (from here on called $V_{\rm rot}$). $V_0$ is the systemic velocity, $\theta$ is the eccentric anomaly and $a_n(r)$ and $b_n(r)$ are the amplitudes at a given radial distance $r$. Therefore, the products of the kinemetry analysis are the kinematic PA and flattening (Q) of the best-fitting ellipses along a set of radii, $V_{\rm rot}$, and higher harmonic terms that describe the variation of the velocities along the ellipse. 

The assumption behind kinemetry is that the motion of gas clouds follow circular orbits in a thin (inclined) disk \citep[e.g.][]{schoenmakers_measuring_1997, wong_search_2004,van_de_ven_kinematic_2010}. Therefore, the dominant kinemetry term, the one next to the $\cos(\theta)$ harmonics, is $V_{\rm rot}=V_{\rm circ}$, which characterises the circular motions in the disk. In case gas clouds are not moving on circular orbits (i.e. due to turbulence or non-settled nature of the disk), the PA and Q describing the ellipse will vary with radius. The variation of PA and Q can be related to the structural changes of the disk, which comprises rings of different orientations and inclinations. A further effect is the non-negligible higher harmonic coefficients beyond $b_1$. A similar principle applies for the stellar orbits in thin disks, but the majority of early-type galaxies actually show regular disk-like kinematics \citep{krajnovic_sauron_2008,krajnovic_atlas3d_2011}. 

\begin{figure}
    \centering
    \includegraphics[width=\columnwidth]{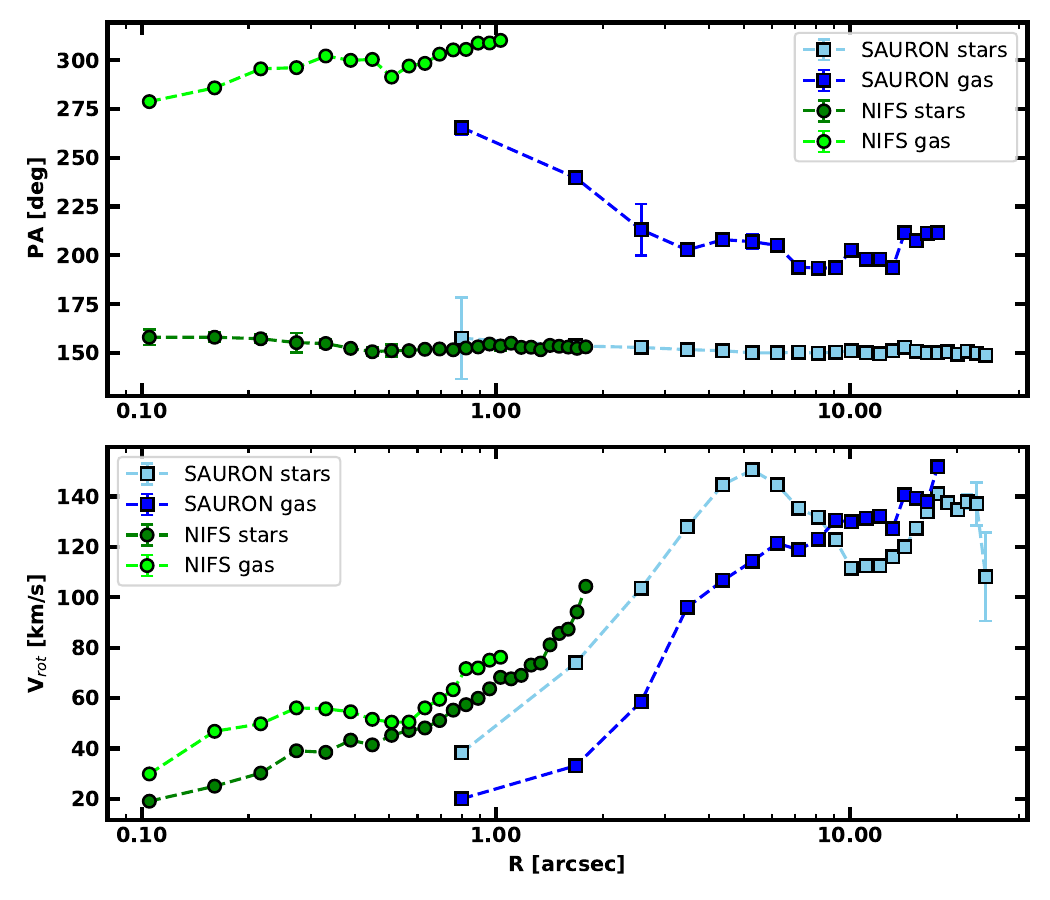}
    \caption{Position angle (top) and rotational velocity (bottom) of the SAURON and NIFS stellar and gas velocity maps as derived with kinemetry. Results for the SAURON kinematics are shown with light-blue and blue squares (stars and gas, respectively), and for the NIFS kinematics with green an lime circles (stars and gas respectively). }
    \label{fig:harm}
\end{figure}

As pointed out previously, the SAURON stellar velocity map of NGC\,4111 was already analysed by \citet{krajnovic_atlas3d_2011}, who concluded that the stellar rotation is indeed ordered and disk-like. Here we repeat this analysis to extend it to the SAURON gas
%MOVE???***, as well as to the NIFS gas and stellar velocity maps, 
to quantify how regular gas motions are. In Fig.~\ref{fig:harm} we show the variation of the position angles of best fitting ellipses, as well as the variation of $V_{\rm rot}$. While running kinemetry on the gas velocity maps, the flattening of the ellipses was poorly constrained. Therefore we imposed a limit on the range of possible Q$>0.98$, making ellipses close to circles. In this way we were still able to recover robust estimates for the PA and $V_{\rm rot}$. Also shown in this figure are the kinemetry results from the NIFS data, to be discussed in Sec.\,\ref{ss:kinemetry_nifs}.

As pointed out above, the visual inspection of the SAURON gas velocity map already shows (see Fig.\,\ref{fig:sauron_kin}), a significant difference in the orientation of gas and stellar kinematics. This difference is now quantified by kinemetry to be about $55\degr$ beyond the inner $\approx$\,3\arcsec. This angle for the gas kinematics does not coincide with any of the principal axes in NGC\,4111 (the kinematic and photometric major axes have the same orientation within errors), and, therefore, the [\ion{O}{iii}] emission line gas is in an unstable configuration. Not only is the gas kinematics misaligned with respect to that of the stars, but there seems to be a continuous twisting of the kinematic position angle towards the nucleus. This change of the gas kinematics position angle suggests the presence of non-circular motions of gas clouds. 

The pattern in the gas velocity dispersion (bottom panel of Fig.\,\ref{fig:sauron_kin}) with increased values in a twisted configuration could be indication of kinematic disturbance in the gas due to the non-circular motions. Alternatively, it could be the result of the superposition of two kinematic components in the gas: the first due to rotation of the gas in the galaxy disk, as for the stars, and the second due to rotation of the gas in the polar ring.

\begin{figure}
    \centering
    \includegraphics[width=\columnwidth]{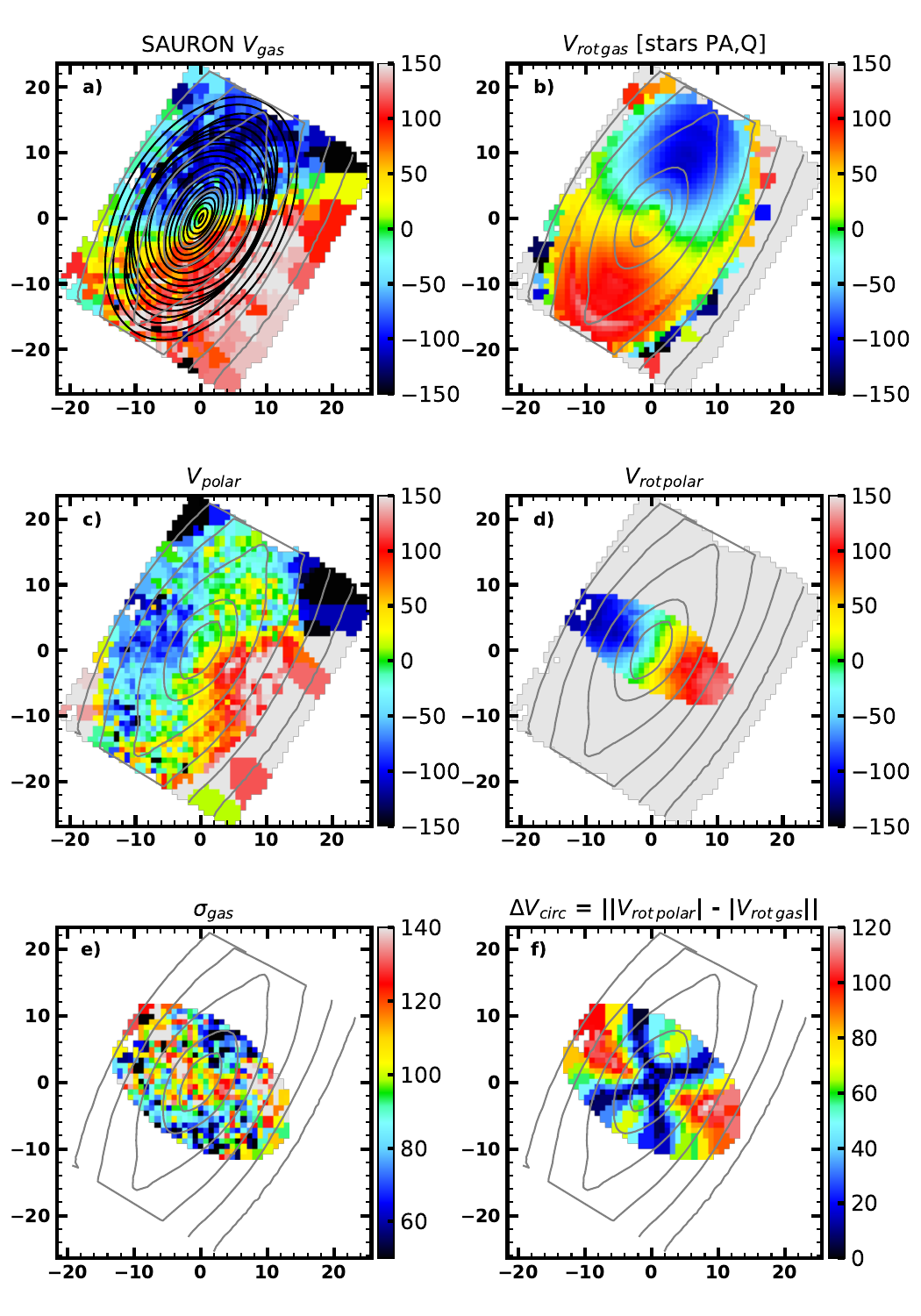}
    \caption{Analysis of the polar motions of the SAURON ionised gas. Panel a): the SAURON gas velocity field as in Fig.~\ref{fig:sauron_kin}, with overploted ellipses used to extract the gas motions in the equatorial plane of the galaxy. Panel b): resulting velocity field of the gas in the equatorial plane of the galaxy. Panel c): the velocity field representing ``polar" motions of the ionised gas, obtained by subtracting velocity field in panel b) (ordered rotation of the gas in the equatorial plane of the galaxy) from the SAURON gas velocity field in panel a).  Panel d): ordered motion of the gas in the polar disk, obtained using kinemetry. Panel e): SAURON gas velocity dispersion map as in Fig.~\ref{fig:sauron_kin}, but limited to the same region as in panel d), where there are data from the polar motions. Panel f): absolute difference between the ordered motions in the equatorial and the polar planes (panels b and d). Note that panels e) and f) have high velocity values in the same regions, providing a qualitative explanation for the unusual structure of the SAURON gas velocity dispersion map. Note also that the colour bars on these panels are not the same, but adjusted to highlight the kinematic features. On all panels grey lines presents the isophotes of the stellar light. }
    \label{fig:resid_new}
\end{figure}

We now investigate the possibility that the SAURON ionised-gas velocity field is indeed due to the superposition of two components, one rotating in the main galaxy (equatorial) plane and the other in the polar ring. Our goal is to check if the features on the SAURON map of the ionised-gas velocity dispersions are consistent with originating from a superposition of gas motions in these two planes. Assuming that we have two gas disks, both characterised by ordered motions, and as the location of high velocities in each of them will be spatially offset, the expectation is that their superposition will effectively be transposed into high velocity dispersion values, offset from the centre, and approximately (but not exactly) along the minor axis of the galaxy.

We used kinemetry to investigate this possibility. The first step was to extract the ordered motions belonging to the gas disk in the equatorial plane. To do this we used the set of best-fitting ellipses from the kinemetric analysis of the stellar velocity field. As the stellar velocity field is very regular, and consistent with axisymmetry \citep{krajnovic_atlas3d_2011}, these ellipses trace circular orbits in the equatorial plane. As Fig.~\ref{fig:harm} shows, the orientation of these ellipses is on average 55\degr offset from the gas motions, as mentioned above. We show the ellipses over-plotted on the gas velocity field in panel a) of Fig.~\ref{fig:resid_new}. The velocity component that is obtained in this way we call $V_{\rm rot\_gas}$, and show it in panel b). Under kinemetry assumptions this component contains the motions of gas clouds in the equatorial plane and on circular orbits. 

The gas component that is not within the equatorial plane can be derived by subtraction, $V_{\rm polar} = V_{\rm gas} - V_{\rm rot\_gas}$, and we show it in panel c). We tentatively call it a ``polar" component as the velocity field is aligned with the minor axis of the galaxy and the observed polar ring. The next step is to run kinemetry on this $V_{\rm polar}$ field. On panel d) of Fig.~\ref{fig:resid_new} we show the ordered motions belonging to this gas component, and call it $V_{\rm rot\_polar}$. This component, under assumptions of the kinemetric analysis, represents the part of the gas that is on circular orbits in the polar disk/ring. 

As mentioned above, our goal is to check if the main features of the gas velocity dispersion field can be reproduced by the superposition of motions in the equatorial and polar planes. We do this by looking at the absolute differences between the velocity fields in panels b) and d), or the difference between the ordered motions in the two planes. The result is shown in panel f) and should be compared with the map in panel e). The two maps are qualitatively similar, showing high and low velocity values at the same locations: nearly along the minor and major axes, respectively. The model velocity dispersion map of panel f) is only good as a simplistic approximation of the observed gas velocity dispersion in the SAURON velocity field of NGC\,4111 (Fig.\,\ref{fig:sauron_kin}), but it highlights its consistency with the presence of two main components in the gas. Therefore, we can conclude that the SAURON gas kinematics is consistent with originating from two components which are located at different planes of the galaxy. 

\subsection{Nuclear scale - NIFS}

Here we discuss the results from the NIFS data, covering the inner $\approx$\,110\,pc radius, at a spatial resolution of $\approx$\,7\,pc.

\subsubsection{Stellar and molecular (H$_2$) gas flux distributions}

The NIFS flux distribution in the continuum is shown in the right panel of Figure \ref{fig:h2ring} integrated within the 2.175-2.200\,$\mu$m wavelength range, revealing a smooth distribution along the galaxy plane from the southeast (SE) to the northwest (NW) of the nucleus.

The H$_2$(1-0) S(1) line flux map, shown in the left panel of Figure \ref{fig:h2ring}, presents an elongated structure running from the north-east (NE) to the south-west (SW) resembling an inclined ring perpendicular to the galaxy plane and that seems not to be centred at the galaxy nucleus (shown as a cross in the figure). The diameter of the ring is $\approx$\,3$^{\prime\prime}$ (210\,pc), oriented perpendicular to the galaxy plane with a projected width of $\approx0\farcs5$ (37\,pc) encircling the nucleus. The H$_2$ flux distribution -- stronger to the NW than to the SE of the nucleus -- suggests that the NW portion corresponds to the near side of the ring.

\subsubsection{Molecular H$_2$ gas mass and excitation}
\label{sec:H2}

\begin{figure}
    \centering
    \includegraphics[width=\columnwidth]{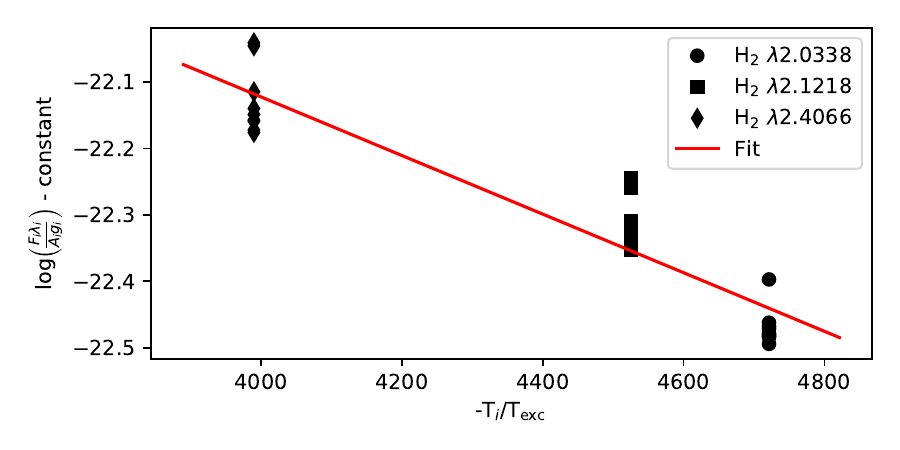}
    \caption{Relation derived using eq. \ref{eq:temp} from \citet{wilman_nature_2005}. The black dot, square and diamond show, respectively, values for the lines H$_2$ $\lambda2.0338\,\mu$m, $\lambda2.1218\,\mu$m and $\lambda2.4066\,\mu$m. The red line is a linear fit to the data, with an angular coefficient $-1/2267$ K$^{-1}$.}
    \label{fig:temp}
\end{figure}

In order to investigate the origin of the H$_2$ emission, we first calculate its temperature. We have looked for other emission lines and found that in some regions of the ring, and in particular at the brightest spot to the SW, at 1\farcs3 (95\,pc) from the nucleus, two other H$_2$ emission lines could be measured: H$_2 \lambda 2.0338\,\mu$m and H$_2 \lambda 2.4066\,\mu$m.

The H$_2$ gas temperature was obtained using the relation from \citet{wilman_nature_2005} applied to the three emission lines above:

\begin{equation}
    \label{eq:temp}
    \text{log} \left(\frac{F_i\lambda_i}{A_ig_i}\right) = \text{constant} - \frac{T_i}{T_\mathrm{exc}}
\end{equation}
\\
\noindent where $F_i$ is the flux from the $i^\mathrm{th}$ $\mathrm{H}_2$ emission line, $\lambda_i$ is its wavelength, $A_i$ is the spontaneous emission coefficient, $g_i$ is the statistical weight of the upper level of the transition and $T_i$ is the level energy expressed as temperature. The expression above is valid for thermal excitation, assuming an \textit{ortho}:\textit{para} abundance ratio of 3:1. From this expression we obtain $T_\mathrm{exc}=2267\,\pm166$\,K for the gas as the inverse of the gradient of the linear fit to the data, as shown in Figure \ref{fig:temp}, adopting the uncertainty from the correlation matrix of the linear regression. $T_\mathrm{exc}$ is the kinetic temperature if the $\mathrm{H}_2$ gas is in thermal equilibrium.

We use the flux distribution in the H$_2 \lambda 2.1218\,\mu$m line to calculate the hot $\mathrm{H}_2$ gas mass according to the expression \citep{scoville_velocity_1982, riffel_mapping_2008}:

\begin{equation}
    \label{eq:gas_hot}
    M_{\mathrm{hot\ H}_2} = \frac{2m_p\ F_{\mathrm{H}_2\lambda2.1218}\ 4\pi D^2}{f_{\nu=1,J=3}\ A_{S(1)}\ h\nu}
\end{equation}
\\
\noindent where $m_p$ is the proton mass, $F_{\mathrm{H}_2\lambda2.1218}$ is the line flux, $D$ is the distance to the galaxy, $f_{\nu=1,J=3}$ is the population fraction and $A_{S(1)}$ is the transition probability. The gas mass is obtained in solar masses. Using T = 2267$\pm166$\,K (obtained above), the population fraction is $f_{\nu=1,J=3} = 1.22 \times 10^{-2}$ and the transition probability is $A_{S(1)} = 3.47 \times 10^{-7}$ s$^{-1}$ \citep{turner_quadrupole_1977, scoville_velocity_1982, riffel_mapping_2008}. Using these parameters, we obtain a mass for the hot molecular gas M$_{\mathrm{hot\ H}_2}=139.48\pm4.38$\,M$_\odot$.

This mass is very small but is only be the ``hot skin" of the H$_2$ total mass, which should be dominated by cold gas. A number of studies have compared H$_2$ masses obtained using the cold CO molecular lines (observed at mm wavelengths) with that of hot H$_2$ observed in the near-IR for the nuclear regions of active galaxies. For example, \citet{dale_warm_2005} obtained ratios in the range $10^5\le$\,M$_{\mathrm{cold\ H}_2}$/M$_{\mathrm{hot\ H}_2}\le10^7$, \citet{muller_sanchez_sinfoni_2006} in the range $10^6\le$\,(M$_{\mathrm{cold\ H}_2}$/M$_{\mathrm{hot\ H}_2})\,\le5\times10^6$, while \citet{mazzalay_molecular_2013}, compiling more data, covering a wider range of luminosities than the previous studies have derived the following relation:

\begin{equation}
    \label{eq:gas_cold}
    M_{\mathrm{cold\ H}_2} \approx 1174 \left(\frac{L_{\mathrm{H}_2\lambda2.1218}}{L_\odot}\right)
\end{equation}
\\

\noindent where L$_{\mathrm{H}_2\lambda2.1218}$ is the luminosity of this H$_2$ line and the standard deviation of the scatter about the factor $\beta = 1174$ is $\approx$\,35\%. Using the above expression we obtain M$_{\mathrm{cold\ H}_2}\approx(1.01\pm0.36)\times10^8$\,M$_\odot$, an order of magnitude larger than the lower limit for the cold gas mass we have obtained from the HST colour map (Sec.\,\ref{ss:hst_images}).

This cold gas mass is comparable to those recently obtained by \citet{schonell_gemini_2019} using the same instrument and methodology for a sample of 6 nearby active galaxies, which are in the range $8\times10^7$M$_\odot\le$\,M$_{\mathrm{cold\ H}_2}\le6\times10^8$\,M$_\odot$.

\begin{figure}
    \centering
    \includegraphics[width=\columnwidth]{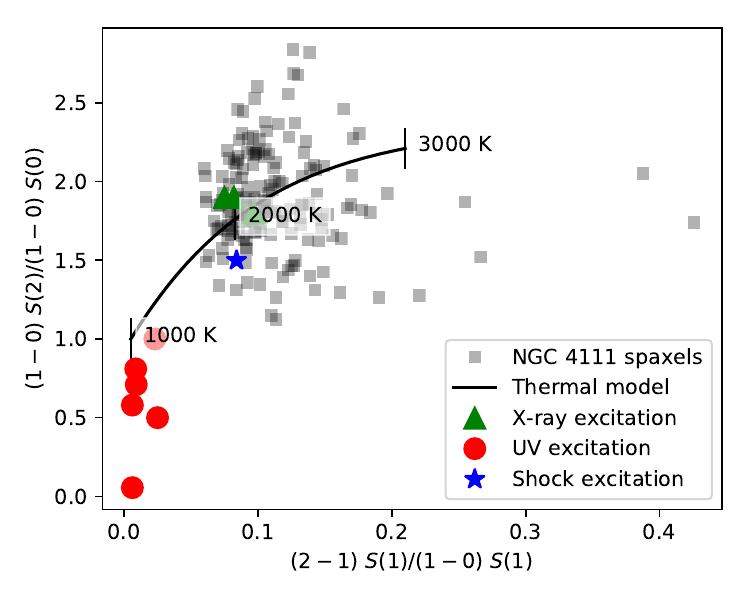}
    \caption{Molecular hydrogen line ratio diagram H$_2$(1-0)S(2)/H$_2$(1-0)S(0) \textit{versus} H$_2$(2-1)S(1)/H$_2$(1-0)S(1) \citep{mouri_molecular_1994}. The gray squares are spaxels from the H$_2$ ring of NGC\,4111. The black curve represents thermal emission models with temperatures in the range 1000-3000\,K. The red circles are the thermal UV excitation models from \citet{sternberg_infrared_1989}, the blue star is the shock heating model from \citet{brand_constancy_1989} and the green triangles are X-ray heating models by \citet{draine_h2_1990}.}
    \label{fig:thermal}
\end{figure}

What physical processes heat the gas to the calculated temperature of $\approx$\,2200\,K? In order to investigate the nature of the gas excitation, we calculated the H$_2$(2-1)S(1)/H$_2$(1-0)S(1) line ratio, as its value can be used to distinguish the excitation mechanism: values of the order of 0.15$\pm$0.15 are consistent with thermal processes and values of the order of 0.55 and larger are caused by UV fluorescence \citep{mouri_molecular_1994, reunanen_near-infrared_2002, rodriguez-ardila_molecular_2004, storchi-bergmann_feeding_2009}. For NGC\,4111, we find a mean value of 0.2 along the H$_2$ ring, suggesting primarily thermal excitation. The heating mechanism is believed to be the nuclear hard X-ray photons, whose source is apparently embedded in the dusty ring \citep{gonzalez-martin_x-ray_2009}. The lack of fluorescence -- that requires the presence of UV photons -- is consistent with the absence of the Br$\gamma$ recombination line in the NIFS spectra, with the UV photons likely being extinguished by the dust surrounding the nucleus and thus not available to ionise the neutral hydrogen. However, the absence of Br$\gamma$ may also be explained by a lack of stellar formation in the nuclear region -- maybe due to heating of the gas by the obscured AGN, despite the existence of available molecular gas in the region.

To further constrain the excitation mechanism, we have used the molecular hydrogen line-ratio diagram from \citet{mouri_molecular_1994}, shown in Figure \ref{fig:thermal}, where we have plotted the ratios H$_2$(1-0)S(2)/H$_2$(1-0)S(0) \textit{versus} H$_2$(2-1)S(1)/H$_2$(1-0)S(1) from the NIFS data as grey squares for the spaxels presenting these lines. This diagram compares line ratios that only take place in \textit{ortho} H$_2$ molecules [the (2-1)S(1) $\lambda$2.2477$\,\mu$m and (1-0)S(1) $\lambda$2.1218$\,\mu$m transitions] with those in \textit{para} H$_2$ molecules [the (1-0)S(2) $\lambda$2.0338$\,\mu$m and (1-0)S(0) $\lambda$2.2235 $\,\mu$m transitions]. We included only SNR$>$2 spaxels in the plot to reduce the scattering caused by noisy spaxels. In the figure, we also include a sequence of the thermal models for temperatures ranging from 1000\,K to 3000\,K as a black curve. In addition, we include predictions by models according to other possible  excitation mechanisms: (1) UV excitation due to fluorescence, shown as red circles \citep{sternberg_infrared_1989}; (2) shock excitation due to outflowing material, e.g. from supernova remnants and/or gas winds,
shown as a blue star \citep{brand_constancy_1989}; (3) X-ray heating from AGN, shown as green triangles \citep{draine_h2_1990}. The model data used in Fig.\,\ref{fig:thermal} follows \citet{mouri_molecular_1994}.

The comparison between our data and the models clearly indicates thermal excitation of H$_2$ with temperatures of about 2000\,K and higher (but lower than 3000\,K), including heating by X-rays and/or shocks.

\subsubsection{Stellar kinematics in the inner 110\,pc}

\begin{figure*}
	\includegraphics[width=\textwidth]{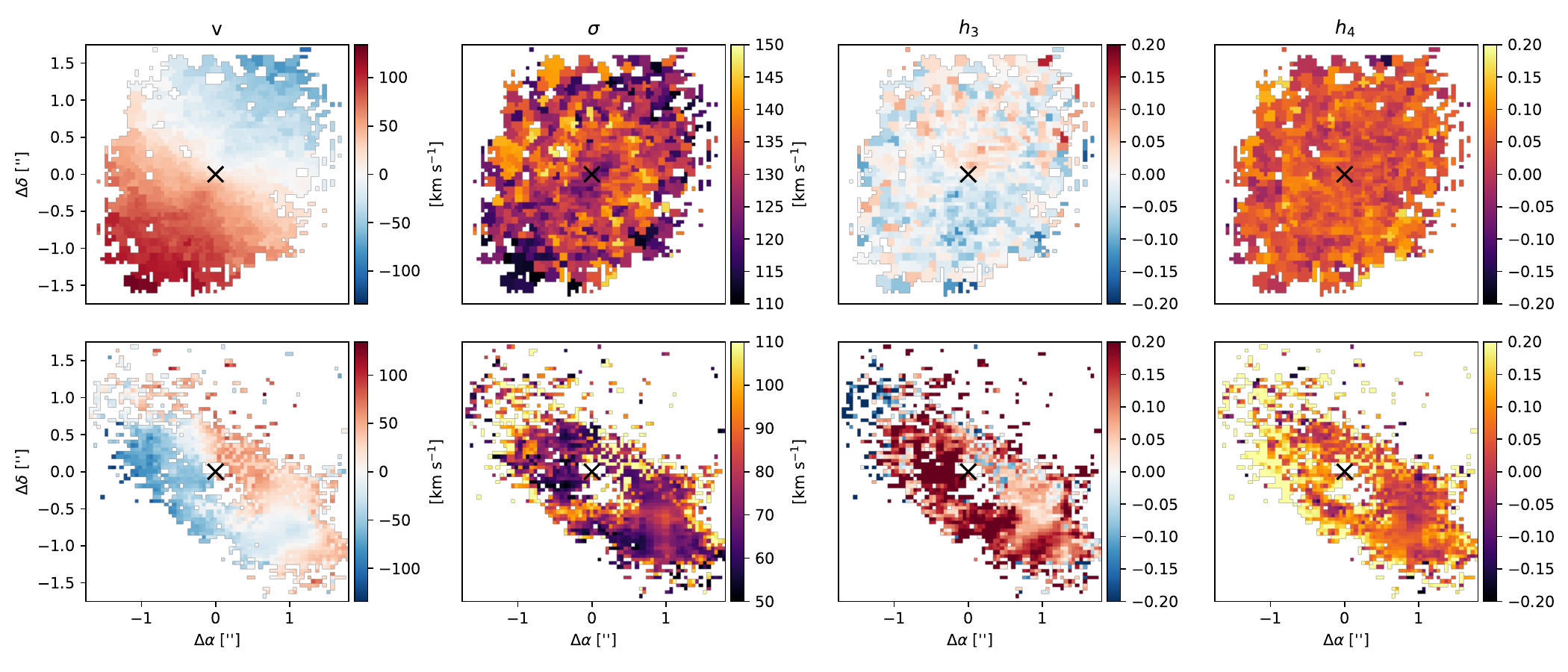}
    \caption{Top: NIFS stellar kinematic maps, from left to right: centroid velocity, velocity dispersion, Gauss-Hermite moments $h_3$ and $h_4$. The respective mean Monte Carlo errors are 4.9\,km\,s$^{-1}$, 6.8\,km\,s$^{-1}$, 0.03 and 0.06. Bottom: As in the top panels, for the molecular gas kinematics (H$_2\ \lambda2.1218\,\mu$m), with mean errors of 17.9\,km\,s$^{-1}$, 39.5\,km\,s$^{-1}$, 0.05 and 0.04, respectively. For error maps see Fig.\ref{fig:nifs_kin_err} of the Appendix. The cross indicates the location of the galaxy nucleus.}
    \label{fig:nifs_kin}
\end{figure*}

The top panels of Figure\,\ref{fig:nifs_kin} show maps of the stellar kinematics, with errors obtained using a Monte Carlo technique, with the corresponding maps shown in  Fig.\,\ref{fig:nifs_kin_err} of the Appendix. These errors decrease outwards, with mean error values for the centroid velocity, velocity dispersion, Gauss-Hermite moments $h_3$ and $h_4$ of 4.9\,km\,s$^{-1}$, 6.8\,km\,s$^{-1}$, 0.03 and 0.06, respectively.

The stellar kinematics of Fig.\,\ref{fig:nifs_kin} reveal a rotation pattern with blueshifts to the NW and redshifts to the SE. A comparison of the top left panel of Fig.\,\ref{fig:nifs_kin} with the top right panel of Fig.\,\ref{fig:sauron_kin} shows that the NIFS stellar velocity field within the inner 110\,pc is similar to that of  the SAURON kinematics at larger scales. The $\sigma$ map shows a partial ring with radius of about 1\arcsec (73.1\,pc) with velocity dispersion values in the range 140--150\,km\,s$^{-1}$ surrounding a small drop to about 120\,km\,s$^{-1}$ within the inner 0\farcs5 ($\sim$37\,pc), similar to the drop seen in the SAURON data at larger scales. This indicates the presence of a dynamically cold structure, such as a disk. Although its origin is not  clear, similar structures have been associated with young to intermediate age stellar population that retains the colder kinematics of the gas from which it has formed and did not reach equilibrium with the bulge velocity field of the older stars yet \citep{riffel_intermediate-age_2010, riffel_intermediate-age_2011}. This would be consistent with a possible scenario in which gas from the polar ring has been captured and is forming new stars in the center of the galaxy.

 The $h_3$-map shows small positive values in the blueshifted side of the velocity field, and, similarly negative values in the redshifted side, indicating that the velocity distribution is slightly asymmetric, showing a "red tail"\, and a "blue tail"\,in the blueshifted and redshifted sides, respectively. The $h_4$-map shows small negative values (broader than a Gaussian) associated to higher velocity dispersion spaxels and positive $h_4$ values (narrower than a Gaussian), for the spaxels with lower velocity dispersion. From Monte Carlo simulations we obtain a mean uncertainty for the velocity, velocity dispersion, $h_3$ and $h_4$ maps of 4.9\,km\,s$^{-1}$, 6.8\,km\,s$^{-1}$, 0.03 and 0.06, respectively.

The measurements of the stellar velocity, velocity dispersion and the higher Gauss-Hermite moments from the NIFS data -- see the $\sigma_{\star_{\NIFS}}$ map in the second panel on the upper row of Figure \ref{fig:nifs_kin} -- will be used in a future work in order to obtain a dynamical model and constrain the SMBH mass M$_\bullet$ (together with those of other targets of our project).

\subsubsection{Molecular gas kinematics}

The maps of the H$_2$ gas kinematic parameters, obtained from the IFSCube fits are shown in the bottom panels of Figure \ref{fig:nifs_kin}, while the corresponding Monte Carlo error maps are shown in Fig.\,\ref{fig:nifs_kin_err} of the Appendix. The mean errors in the parameters $v$, $\sigma$, $h_3$ and $h_4$, are, respectively 17.9\,km\,s$^{-1}$, 39.5\,km\,s$^{-1}$, 0.05 and 0.04.

The H$_2$ velocity field shows redshifts to the NW and blueshifts to the SE, thus opposite to the observed velocities of the stellar velocity field. This kinematics is also distinct from the [\ion{O}{iii}] kinematics, that present blueshifts to the S-SW and redshifts to the N-NE. Nevertheless, we point out that the [\ion{O}{iii}] emission line probes the ionised gas kinematics on kpc scales and the H$_2$ emission probes the molecular gas kinematics at 10-100 pc scales. The velocity dispersion is roughly $\sim$70\,km\,s$^{-1}$ all along the H$_2$ ring, with some higher values from the NW to the NE part of the ring. The $h_3$-map shows the same kind of correlation with the velocity field as shown by the stellar kinematics maps: positive $h_3$ values in the blueshifted side of the ring -- a "red tail"\, on the profiles -- and negative values for the redshifted side of the ring --  a "blue tail"\, on the profiles. The positive $h_3$ values are nevertheless much higher than the negative values, indicating the presence of red wings in most regions of the ring. The $h_4$ values are predominantly positive, presenting a similar relation to that observed for the stellar velocity field, where negative values are in general associated with higher velocity dispersion and positive values with lower velocity dispersion. From Monte Carlo simulations, we obtain a mean uncertainty for the velocity, velocity dispersion, $h_3$ and $h_4$ maps of 17.9\,km\,s$^{-1}$, 39.5\,km\,s$^{-1}$, 0.05 and 0.04, respectively, which are quite higher than the uncertainties in the stellar kinematics. We present the Monte Carlo error maps for the NIFS stellar and molecular gas kinematics in Figure\,\ref{fig:nifs_kin_err} of the Appendix.

As pointed out above, the H$_2$ gas kinematics (bottom panels of Fig.\,\ref{fig:nifs_kin}) suggest counter-rotation of the molecular ring relative to the stellar kinematics  -- but the kinemetry analysis of Section \ref{ss:kinemetry_nifs} below reveal an angle between the stellar and gas apparent rotation varying in the range 120$^{\circ}$--160$^{\circ}$, thus smaller than 180$^{\circ}$, suggesting another interpretation. As the apparent near side of the ring -- that we hypothesise is the NW because it is brighter than the SE -- shows redshifts and the back (far) side shows blueshifts, an alternative interpretation is that the ring of molecular gas could be falling towards the galaxy center.

\begin{figure*}
    \centering
    \includegraphics[width=\textwidth]{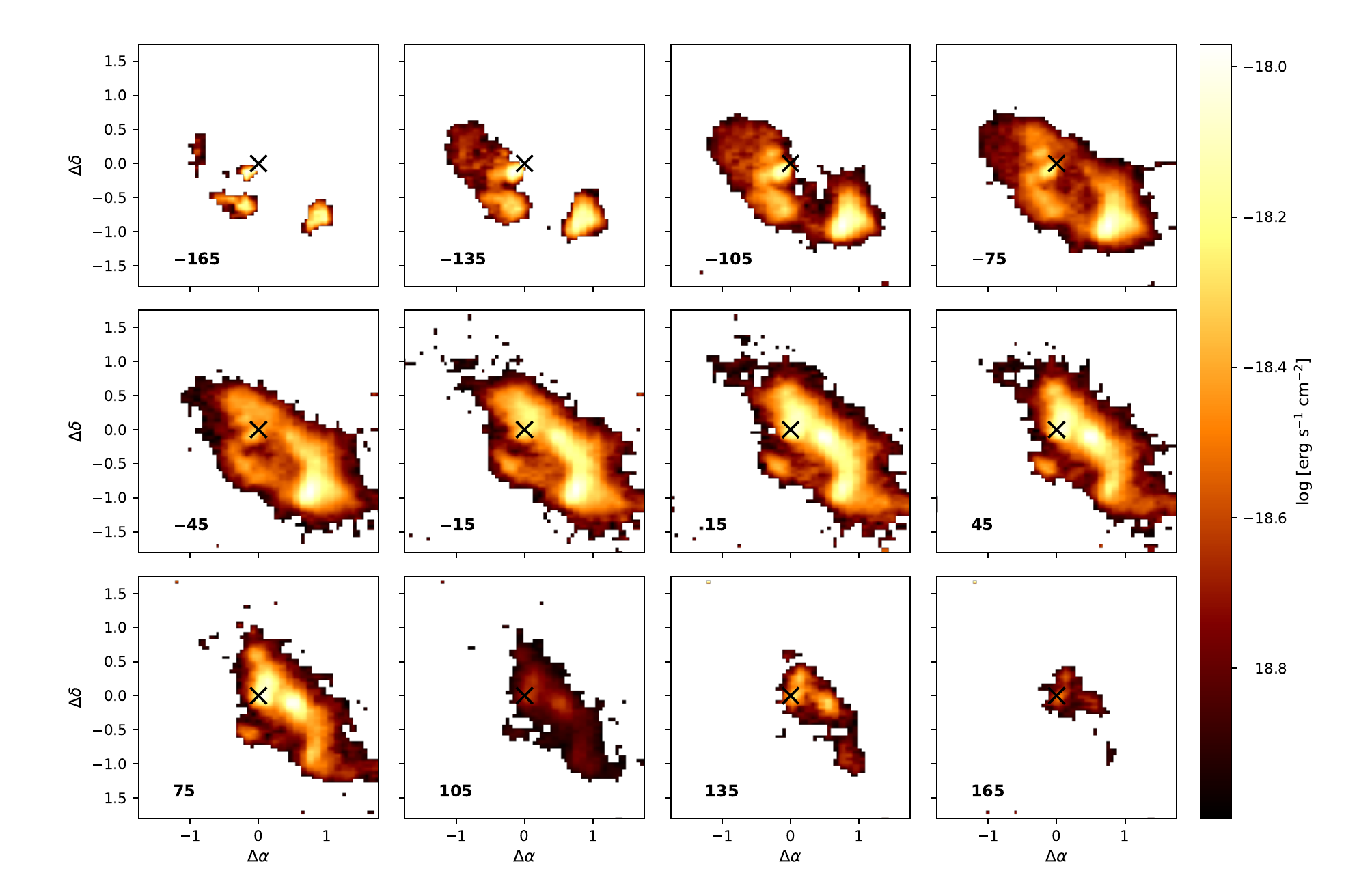}
    \caption{Channel maps of the H$_2 \lambda2.1218\,\mu$m emission line, centered at the velocities listed in the bottom left of each panel in units km\,s$^{-1}$. The cross indicates the location of the galaxy nucleus.}
    \label{fig:cm}
\end{figure*}

\noindent{\it H$_2$ Channel Maps --}
The higher-order moment maps of Fig.\,\ref{fig:nifs_kin} indicate some complexity in the H$_2$ kinematics that is better illustrated via the channel maps shown in Fig.\,\ref{fig:cm}, obtained with IFSCube.

In these maps, the H$_2$(1-0)\,S(1) line profiles have been sliced in velocity bins centred on velocities ranging from -165\,km s$^{-1}$ to 165\,km\,s$^{-1}$ with a 30\,km\,s$^{-1}$ channel width. They show that the fastest blueshifts are observed in 3 compact regions: the smallest is just $\approx$0\farcs2 (15\,pc) to the SE of the nucleus, the second is 0$\farcs5$ south of the nucleus and the last at 1\farcs3 (95\,pc) to the SW, corresponding to the brightest spot of the molecular ring (seen in the left panel of Figure \ref{fig:h2ring} marked with a blue \textit{B}). In the -135\,km\,s$^{-1}$ to -75\,km\,s$^{-1}$ velocity range, the 3 aforementioned regions gradually enlarge and begin to reveal the SE half of the annular structure. The bright spot 1$^{\prime\prime}$ SW of the nucleus seems to keep its flux up to -15\,\kms. The -75\,km\,s$^{-1}$ to 45\,km\,s$^{-1}$ channels show that the flux gradually increases to the NW of the nucleus and fades from the SE, with only emission in a knot to the south and emission to the SW of the nucleus still being detected. In the 75\,km\,s$^{-1}$ to 165\,km\,s$^{-1}$ range, the flux in each channel gradually diminishes, increasing again at 135 km s$^{-1}$. The fastest redshifted component at 165 km s$^{-1}$ is located within $\approx 0\farcs5$ (36\,pc) of the nucleus, towards the NW, opposite to the fastest blueshifted region just to the SE of the nucleus. We interpret this change in orientation as follows: if the NW is indeed the near side of the ring, this velocity pattern does indicate infall towards the center of the galaxy. We do not see indication of rotation along the ring, as, if this were the case, we should observe the highest velocities at the SW and NE tips of the ring, that should also be opposite to each other. We do not see this; the highest velocities are observed to the SE and NW. The channel maps further show that the highest velocity regions tend to be closer to the nucleus than the lower velocity regions, suggesting an inward acceleration, also consistent with an inflow scenario.

\begin{figure}
    \centering
    \includegraphics[width=\columnwidth]{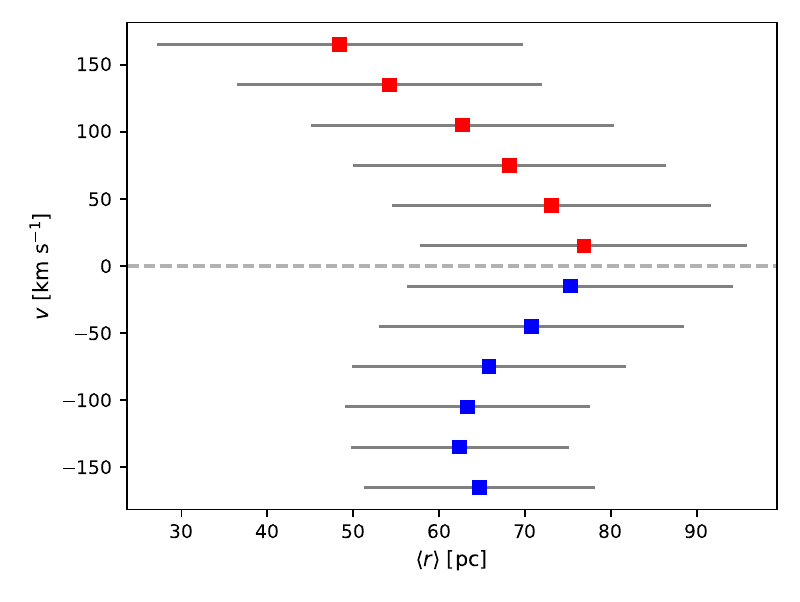}
    \caption{Average projected radial distance $\langle r \rangle$ for each kinematic channel of the H$_2$ emitting gas from Fig.~\ref{fig:cm}, including the RMS variation of the distances in each channel as a gray bar. In red (blue) we indicate the redshifted (blueshifted) channels.}
    \label{fig:cd}
\end{figure}

To further evidence the signature of inflow, we calculated the average projected radial distance $\langle$r$\rangle$ of the H$_2$ gas for each kinematic channel. We present the channel velocity values as a function of these radii in Figure~\ref{fig:cd}. As expected, the mean radial distances decrease as the velocities increase, supporting acceleration of the molecular gas towards the nucleus in an inflow.

\subsubsection{Kinemetric analysis}
\label{ss:kinemetry_nifs}

We have also applied kinemetry to investigate the NIFS stellar and molecular gas kinematics, as we have done to investigate the SAURON kinematics.

The results from the kinemetry analysis for the stellar and H$_2$ gas velocity fields from the NIFS data -- major axis $PA$ and rotation velocities $V_{rot}$ are shown, together with those for the SAURON data in the two panels of Fig.~\ref{fig:harm}.

The NIFS stellar velocities rise more steeply than the SAURON velocities (as expected given the sharper PSF of NIFS), but otherwise are fully consistent. One other observation is worth remarking: while H$_2$ and stellar velocities from the NIFS data have similarly steep rise and amplitudes, molecular gas being marginally faster \citep[expected, e.g.][]{young_structure_2008}, the ionised gas in the SAURON data is significantly slower in the central region. This is a clear indication that its clouds are neither in a thin disk configuration nor are on circular orbits. 

\begin{figure*}
    \centering
    \includegraphics[width=\textwidth]{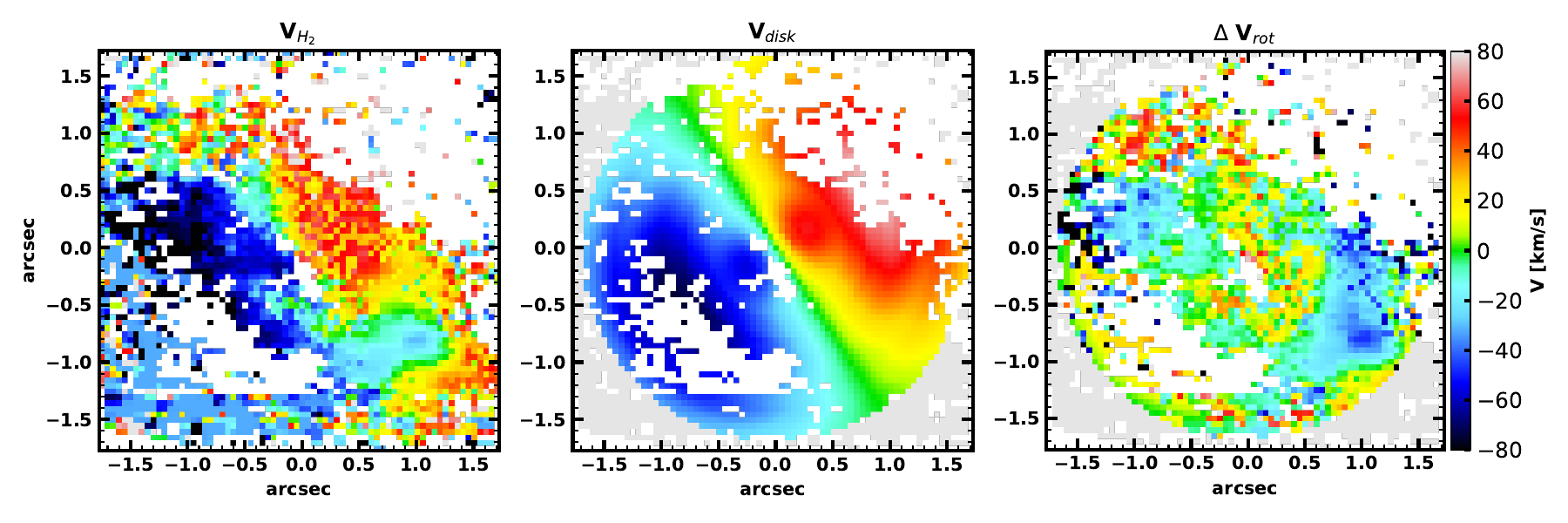}
    \caption{NIFS H$_2$ kinematics (left), its kinemetry fitting (center) and the residual between the two (right), showing evidence of non-circular components in the velocity field.}
    \label{fig:h2_kin}
\end{figure*}

The top panel of Fig.~\ref{fig:harm} shows that the H$_2$ kinemetry reveals a misalignment between the H$_2$ and stellar kinematics of $\approx 155\degr$, indicating that, besides the ionised gas on large scales being misaligned relative to the stellar kinematics, the gas in the central region is also in an unstable configuration, and that gas clouds are not moving on purely circular orbits. We show this in the three panels of Fig.~\ref{fig:h2_kin}. The first panel shows the same H$_2$ velocity field of Fig~\ref{fig:nifs_kin}, which is followed by the kinemetric reconstruction of a ``circular" velocity map, $V_{\rm disk}$. This map is reconstructed using the $V_{\rm rot}$ of the molecular gas (as shown in Fig~\ref{fig:harm}), but assuming that the orientation of the field is constant: we fixed it to the median value of $305\degr$ (top panel of Fig~\ref{fig:harm}). As the H$_2$ gas shows some variation in the PA and large values of the other harmonic coefficients, $V_{\rm disk} \ne V_{\rm circ}$. Nevertheless, it can be used as first order approximation to gauge the contribution
%quantify the contribution 
of non-circular motions in the molecular gas kinematics, shown on the residuals in the last panel of Fig.~\ref{fig:h2_kin}.

\section{Scenario}

In the present section, we use the results from the large scale gas distribution (HST), excitation and kinematics (SAURON) as compared to those of the nuclear scale (NIFS) to propose a scenario for the physical processes occurring within the inner kpcs of NGC\,4111.

\subsection{Comparison between the large and nuclear scale kinematics}

The stellar velocity maps of the large-scale SAURON and nuclear scale NIFS data -- characterized via kinemetry in Fig.\,\ref{fig:harm} are remarkably similar: they share the same position angle and flattening (not shown), and have minimal residuals in higher order harmonic coefficients (also not shown).

\begin{figure*}
    \centering
    \includegraphics[width=\textwidth]{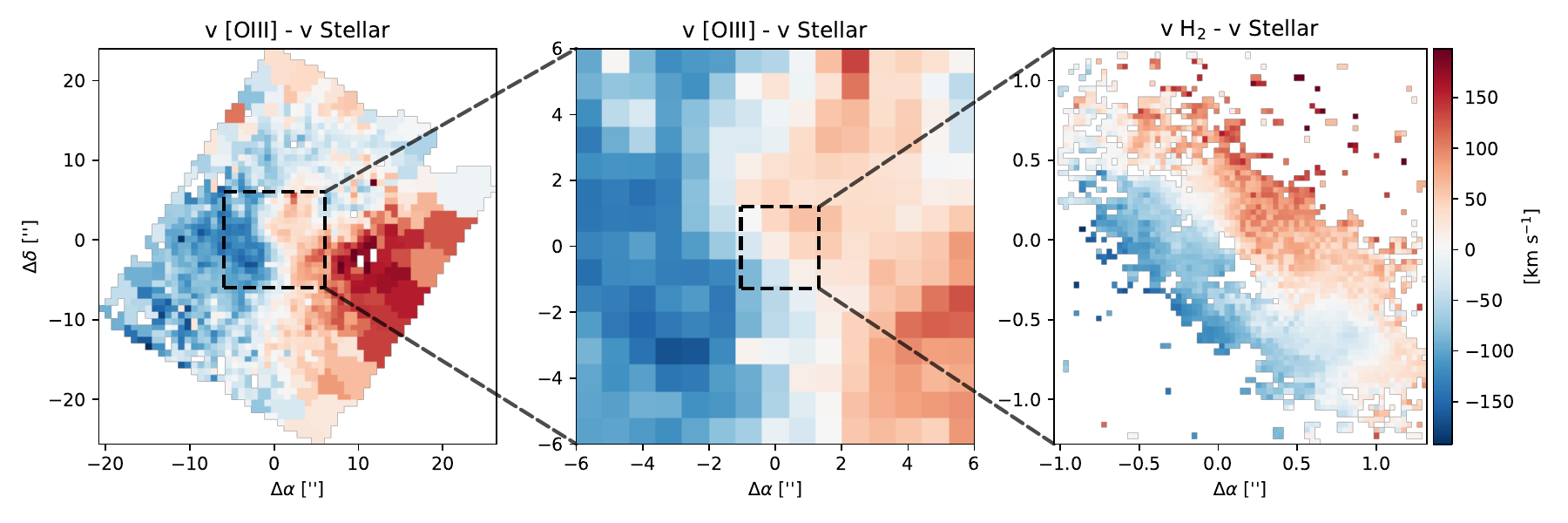}
    \caption{Comparison between the kpc-scale (SAURON) v[\ion{O}{iii}] and the 100\,pc scale (NIFS) H$_2$ gas velocity fields after the subtraction of the stellar velocity field. In the left panel, the dashed square shows the FoV from the central panel; in the central panel, the dashed square shows the NIFS FoV over the zoomed-in velocity field of the left panel.}
    \label{fig:oiii_h2}
\end{figure*}

Nevertheless, the large-scale [\ion{O}{iii}] gas kinematics and the nuclear-scale H$_2$ kinematics look distinct, but a closer look reveals that they seem to be physically connected. This connection between the large-scale ionised gas and the nuclear-scale molecular gas is suggestively shown in Figs.\,\ref{fig:oiii_h2} and \ref{fig:nifs_zoom}.

Fig.\,\ref{fig:oiii_h2} shows a comparison between the residuals of the [\ion{O}{iii}] and H$_2$ velocity fields relative to the stellar velocity field. Inspection of the [\ion{O}{iii}]  residuals within the NIFS FoV suggest blueshifts to the SE and redshifts to the NW, as observed in the H$_2$ residuals, suggesting that the large scale ionised gas kinematics connect with the nuclear scale H$_2$ gas kinematics. 

We now consider the hypothesis discussed in Section \ref{ss:kinemetry} in which the SAURON ionised-gas velocity field is due to the superposition of two kinematic components, one rotating in the main galaxy (equatorial) plane and the other in the polar ring. In the top panel of  Fig.\,\ref{fig:nifs_zoom} we have used the results from the kinemetry of the polar ring (Fig.\,\ref{fig:h2_kin}) to obtain residuals between the observed velocities and its reconstructed circular velocity field. The residuals suggest the presence of a pattern in the shape of spiral arms, outlined by the black lines in the figure. This spiral configuration is suggestive of inflows, as observed in previous kinematic studies \citep[e.g.][and references therein]{storchi-bergmann_observational_2019}. The small square in the top panel shows the NIFS FoV, with the NIFS H$_2$ velocity field shown in the bottom panel. Although the angular resolution of the SAURON data is much lower that that of the NIFS data, a careful inspection of the residual velocity field of the top panel shows that, within the NIFS FoV (black square) there is a hint of residual redshifts towards the NW and residual blueshifts to the SE, suggesting again a connection between the large-scale ionised gas velocity field and the nuclear scale H$_2$ velocity field, which also shows redshifts to the NW and blueshifts to the SE.

The above results, in which the large scale [\ion{O}{iii}] gas kinematics seems to connect with the nuclear scale H$_2$ gas kinematics implies that, if we can conclude that the H$_2$ kinematics indicates inflow, this inflow seems to be present also in the larger scale gas, or at least originate in the large scale gas.

\subsubsection{The Scenario}

We propose the following scenario to explain the observations from the inner $\approx$1.5\,kpc radius probed by the HST and SAURON\ data down to the inner $\approx$\,110-10\,pc radii probed by the NIFS data. The dusty polar ring probably originates from the capture of a dwarf galaxy by NGC\,4111, consistent with the estimated lower limit of the gas mass of $\approx$\,10$^7$\,M$_\odot$ obtained from the extinction map of the 450\,pc ring. The ionised gas seen in the SAURON data is present in both the galaxy disk and the polar ring structures. The increase in its velocity dispersion in the region of the polar ring can be attributed to the superposition of two kinematic components: rotation in the galaxy plane and orbital motion in the polar ring. This orbital motion shows a disturbed rotation pattern, revealed by the kinemetry analysis, such that the residual motion relative to the rotation suggests the presence of a spiral-like structure that seems to connect with the H$_2$ kinematics within the inner $\approx$\,110\,pc, as suggested by the two panels of Fig.~\ref{fig:oiii_h2}.

\begin{figure}
    \centering
    \includegraphics[width=\columnwidth]{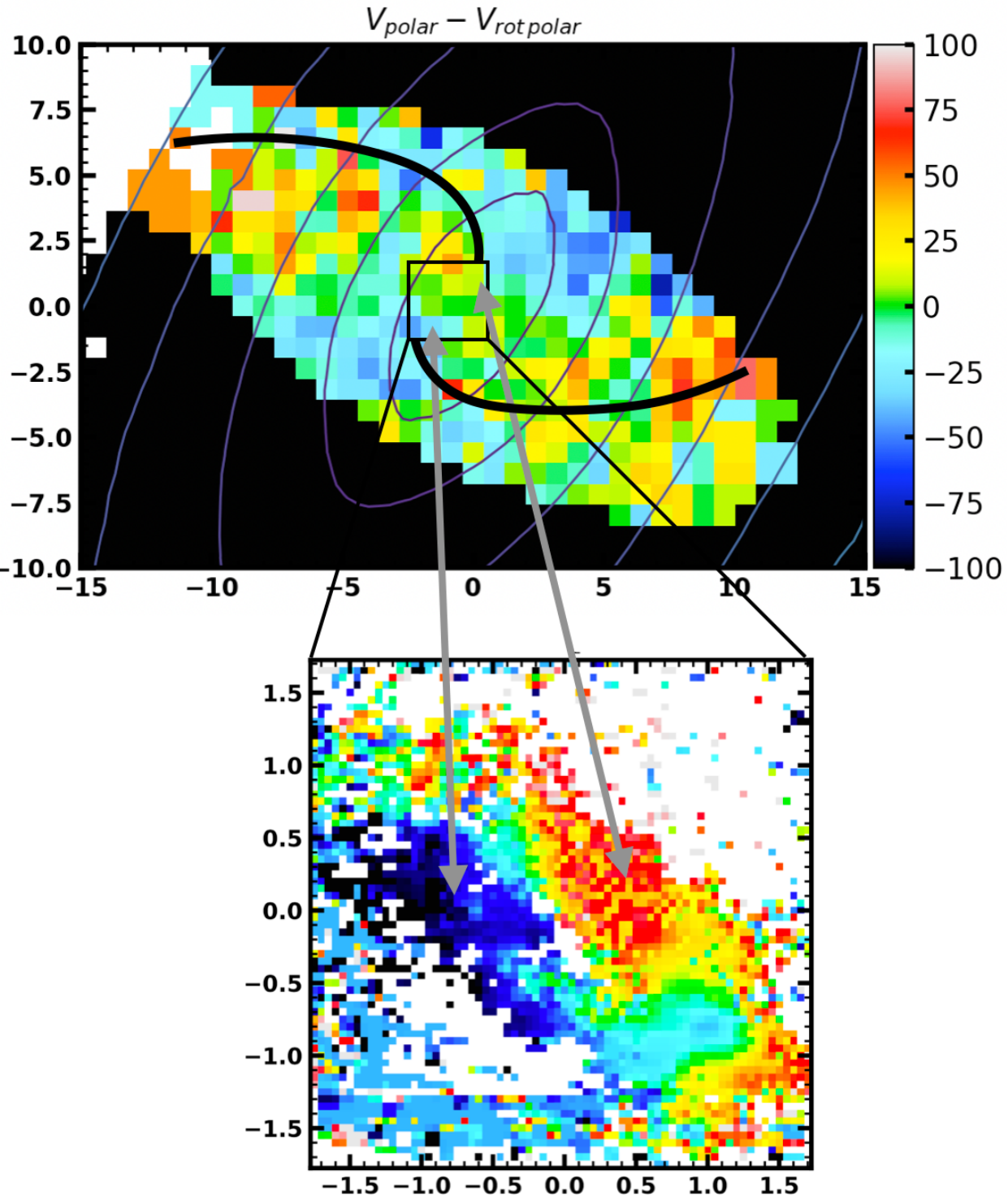}
    \caption{Top: residual map between the ionised gas polar ring velocities and its reconstructed circular velocity field from kinemetry, with the black lines delineating an apparent spiral structure. The small square shows the 3\arcsec$\times$3\arcsec\ NIFS FoV. Bottom: the NIFS H$_2$ velocity field showing redshifts to the NW and blueshifts to the SE that apparently connect with the residuals of the SAURON gas kinematics.}
    \label{fig:nifs_zoom}
\end{figure}

The residual inspiral in NGC\,4111, seen in Fig.~\ref{fig:nifs_zoom}, resembles nuclear spirals (on 100\,pc scales) associated to inward motion of gas found in previous studies of the AGNIFS (AGN Integral Field Spectroscopy) group, such as those in the nearby active galaxies NGC\,7213 \citep{schnorr-muller_gas_2014}, NGC\,1667 \citep{schnorr-muller_gas_2017} and NGC\,2110 \citep{diniz_outflows_2019}, and also the nuclear spirals found in NGC 1097 \citep{davies_stellar_2009, van_de_ven_kinematic_2010, fathi_alma_2013}. Other observational \citep[e.g.][]{combes_alma_2014} and theoretical \citep{kim_nuclear_2017} studies also support the presence of inflows along nuclear spirals in nearby active galaxies.

The H$_2$ ring kinematics -- that seems to connect with the above spirals, suggests inward motion: assuming that the brighter NW side is the near and the fainter SE side is the far, the blueshifts observed in the far side and redshifts observed in the near side indicate inflows. The channel maps of Fig.\,\ref{fig:cm} and the fact that the average radius of the channel maps decrease as the velocity increases, support such inflows, %with some of the regions with the highest velocities observed closer to the galaxy nucleus, 
suggesting, in addition, inward acceleration.

The inflowing gas may be settling in the galaxy plane and forming new stars there, and could be the cause of the observed sigma drop in the NIFS and SAURON data stellar kinematics. Both $\sigma$-drop structures could be part of the same gas-rich merger event, building the large-scale and small-scale disks and gas structures as the merger evolves. At least part of the inflowing gas seems to be reaching the nucleus, triggering a low-luminosity AGN that can be observed in X-rays. This emission -- possibly combined with shocks in the region -- is the source of excitation of the H$_2$, as discussed in Section\,\ref{sec:H2}.

Having concluded that the observed gas kinematics indicate inflows towards the nucleus, we have used the NIFS data to estimate the mass inflow rate and compare it to the mass accretion rate to the obscured AGN. For this, we need first to obtain an estimate for the SMBH mass as well as the AGN accretion rate and Eddington ratio.

\subsection{SMBH mass and Eddington ratio}

We use the integrated $\sigma_\star$ value within one effective radius $\sigma_{\star_{SAURON}} = 163.3 \pm 8.1$\,km\,s$^{-1}$, as previously determined from the SAURON data by \citet{cappellari_atlas3d_2013} in order to estimate M$_\bullet$ with the empirical $M_\bullet - \sigma_\star$ relation from \citet{van_den_bosch_unification_2016}:

\begin{equation}
    \log\left(\frac{M_\bullet}{M_\odot}\right) = (8.32 \pm 0.04) + (5.35 \pm 0.23)\log\left(\frac{\sigma_\star}{200\ \text{km s}^{-1}}\right)
\end{equation}
\\
\noindent For $\sigma_\star=\sigma_{\star_{SAURON}}$, we obtain M$_\bullet = (7.06^{+2.62}_{-1.79}) \times 10^7$\,M$_\odot$. This value is consistent with a previous estimate by \citet{nyland_atlas3d_2016} of M$_\bullet = 6.31 \times 10^7$\,M$_\odot$, obtained with the $M_\bullet - \sigma_\star$ relation from \citet{mcconnell_revisiting_2013}.

The bolometric luminosity $L_\mathrm{bol}$ was obtained from the X-ray luminosity in the 0.5 - 10.0 keV range \citep{gonzalez-martin_x-ray_2009}, assuming that this luminosity represents 8\% of the total luminosity of the AGN \citep{nemmen_radiatively_2006}. This allows us to calculate the Eddington ratio $L_\mathrm{bol}/L_\mathrm{Edd} = 1.5^{+0.5}_{-0.4}\times10^{-4}$ for NGC\,4111, using our determined value of the SMBH mass above. The low Eddington ratio suggests that, if there is an active nucleus in NGC\,4111, as indicated from the emitted X-rays, the accretion regime is that of a Radiatively Inefficient Accretion Flow (RIAF).

\subsection{Mass inflow rate and the mass accretion rate}

We have estimated the mass inflow rate $\dot{m}_\mathrm{inf}$ using the channel maps from Fig.~\ref{fig:cm} and the following expression:

\begin{equation}
  \dot{m}_\mathrm{inf} = \frac{mv}{\langle r \rangle}  
\end{equation}

\noindent where $\langle r \rangle$ is the average radial distance of H$_2$ in each channel (Fig.\,\ref{fig:cd}), $v$ is the channel velocity and $m$ is the integrated gas mass from each channel, determined using eq.~\ref{eq:gas_hot} for the hot molecular gas and eq.~\ref{eq:gas_cold} for the cold.

One important question is: What fraction of the gas is actually inflowing? In our previous studies cited above, as well as suggested in Figs.\,\ref{fig:h2_kin} and \ref{fig:nifs_zoom}, the residuals from rotation orbital velocity are usually observed in spiral structures covering  $\approx$10\%\ of the observed velocity field. We have then tentatively considered this fraction of the gas in each channel that is actually inflowing.

We first considered only the hot phase of the gas in this calculation and obtain a mass inflow rate of only  $7.8^{+3.4}_{-1.7}\times 10^{-4}$\,M$_\odot$ yr$^{-1}$. 
Nevertheless, as discussed above, the hot phase of the gas is probably only ``the heated skin" of a much larger gas reservoir, dominated by the cold gas.

Considering now the cold component and that gas from all channels is inflowing, we determine an upper limit for the mass inflow rate of 56.9$^{+24.1}_{-12.5}$\,M$_\odot$ yr$^{-1}$.

We then considered that only the 2 highest velocity channels (-165, -135, 135 and 165 km\,s$^{-1}$) trace the inflows, with the lower velocity gas tracing gas orbiting in the galaxy plane. The resulting mass inflow rate is  33.9$^{+16.3}_{-7.9}$\,M$_\odot$ yr$^{-1}$.

In order to obtain still another estimate, instead of using eq.\,\ref{eq:gas_cold} to estimate the mass of cold gas, we used the value of hot molecular gas obtained with eq.~\ref{eq:gas_hot} and the lower limit of the ratio between the cold and hot molecular gas masses from \citet{dale_warm_2005} of $10^5$. This way we obtain a lower limit of 7.9$^{+3.3}_{-1.7}$ M$_\odot$ yr$^{-1}$ for the mass inflow rate.

The mass accretion rate $\dot{m}_\mathrm{acc}$ was determined using the expression:

\begin{equation}
    \dot{m}_\mathrm{acc} = \frac{L_\mathrm{bol}}{\eta c^{2}}
\end{equation}

\noindent using the previously calculated $L_\mathrm{bol}$ and assuming an accretion efficiency $\eta = 0.01$, justified by the low Eddington ratio for this AGN  \citep{nemmen_radiatively_2006}, the resulting mass accretion rate is 0.002 M$_\odot$ yr$^{-1}$, which supports the presesnce of a RIAF in the nucleus of NGC\,4111.

The above accretion rate is larger than the mass inflow rate in hot molecular gas, but is much lower than the estimated mass inflow rate in cold molecular gas, that should dominate the mass of molecular gas, as previously pointed out. This indicates that cold molecular gas could be accumulating in the nuclear region and is enough to trigger new episodes of star formation and stronger nuclear activity in the near future. At a mass-inflow rate of $\approx$5--10\,M$_\odot$ yr$^{-1}$, the existing cold gas reservoir of $\approx$\,10$^{8}$M$_\odot$ should keep this process going for the next 10$^7$ years.

\section{Conclusions}

We have analysed galactic scale HST images of the nearby S0 polar-ring galaxy NGC\,4111 through the F475W and F160W filters together with SAURON optical IFS of the inner 1.5\,kpc radius and nuclear scale K-band Gemini NIFS AO-assisted IFS of the inner 110\,pc radius. Our main conclusions are:

\begin{itemize}
    \item The HST images reveal the dusty polar ring extending by $\sim$\,450\,pc perpendicular to the plane of the galaxy. From the colour map F475W-F160W we obtain an extinction map and a lower limit for the cold gas mass in the ring as M$_{\text{gas}} = 9.8 \times 10^{6}$ M$_\odot$. This mass is consistent with an origin from a captured dwarf galaxy. 
    
    \item Ionised gas emission is seen in the SAURON data, with line ratio values $1\le$[\ion{O}{iii}]$\lambda4959,5007\,\AA/\mathrm{H\beta}\le$3 supporting the LINER classification of previous studies.
    
    \item The NIFS data show a hot molecular gas ring with diameter of 220\,pc – embedded in the dusty polar ring above – with a calculated temperature of 2267$\pm$166\,K and mass M$_{\rm {hot\,H_2}}$=139.5$\pm$ 4.4M$_\odot$. This hot gas mass corresponds to an estimated cold gas mass of M$_{\rm{cold\,H_2}}\approx$\,10$^8$\,M$_\odot$ within the inner 110\,pc (radius) which is consistent with the lower limit for the gas mass obtained from the HST images and further suggests an extragalactic origin for this gas.
    
    \item The excitation of the hot H$_2$ is thermal, most probably due to X-ray heating, supporting the presence of an AGN at the nucleus. Although shocks could also contribute to the gas excitation, the low velocity dispersion ($\sim$\,60\,km\,s$^{-1}$) does not support a major contribution of shocks to the H$_2$ excitation.

    \item The large scale (1.5\,kpc radius) SAURON and smaller scale  (110\,pc radius) NIFS stellar velocity distributions show the same rotation pattern, as supported also by kinemetry analysis.
    
    \item Both the large and small scale stellar velocity dispersion $\sigma_\star$ show drops in their values along the galaxy major axis, within radii of $\approx$\,350\,pc and $\approx$\,35\,pc, respectively. This is probably associated to recent capture of gas (on both scales) that settled in a disk and formed new stars that have still kept their colder gas kinematics as compared to the surrounding bulge. The presence of younger stars in the region is supported by previous studies of the stellar population.

    \item Both the 1.5\,kpc scale [\ion{O}{iii}] and the 110\,pc scale hot H$_2$ velocity fields are distinct from the stellar one, indicating that the gas in both ionised and molecular phases is not in a stable configuration. 
    
    \item The [\ion{O}{iii}] gas kinematics show two components: (1) rotation similar to that of the stars in the galaxy plane; (2) disturbed rotation along the dusty polar ring. Velocity residuals relative to the rotation component in the ring show an spiral pattern suggestive of inflows that seem to connect to the H$_2$ velocity field in the 110\,pc ring.
    
    \item The H$_2$ velocity field in the 110\,pc ring shows blueshifts in the apparent far side and redshifts in the apparent near side, as well increasing velocity towards the inner regions in channel maps, indicating inflows towards the nucleus.
    
    \item We have estimated the mass inflow rate towards the nucleus from the H$_2$ channel maps and the expected corresponding cold molecular gas mass, finding values of 5--10\,M$_\odot$\,yr$^{-1}$. Such values are much larger than the accretion rate to the AGN and enough to feed new episodes of star formation within the inner few pc--100\,pc considering the available gas reservoir of $\approx\,10^8$\,M$_\odot$.
    
\end{itemize}

We conclude that our data supports a scenario in which a dwarf, gas-rich galaxy has been recently captured by NGC\,4111 giving origin to its polar ring. Part of this gas is flowing in and possibly forming new stars while another part of it seems also to have just reached the nuclear SMBH triggering a so-far low-luminosity AGN. This AGN is already emitting X-rays that are heating the gas and causing thermal excitation of the H$_2$ molecule. However, as we do not see Br$\gamma$ in our K-band data, UV photons are not reaching the observed gas, which suggests that either new stars have not begun to form and/or these stars and the AGN are still buried in the dust rich environment, supporting that the AGN has only recently been triggered.

\section*{Acknowledgements}

We thank the anonymous referee that helped us in improving our paper.

This research is based on observations obtained at the international Gemini Observatory, a program of NSF’s NOIRLab, which is managed by the Association of Universities for Research in Astronomy (AURA) under a cooperative agreement with the National Science Foundation. on behalf of the Gemini Observatory partnership: the National Science Foundation (United States), National Research Council (Canada), Agencia Nacional de Investigaci\'{o}n y Desarrollo (Chile), Ministerio de Ciencia, Tecnolog\'{i}a e Innovaci\'{o}n (Argentina), Minist\'{e}rio da Ci\^{e}ncia, Tecnologia, Inova\c{c}\~{o}es e Comunica\c{c}\~{o}es (Brazil), and Korea Astronomy and Space Science Institute (Republic of Korea).

This research is based on observations made with the NASA/ESA Hubble Space Telescope obtained from the Space Telescope Science Institute, which is operated by the Association of Universities for Research in Astronomy, Inc., under NASA contract NAS 5–26555. These observations are associated with program GO-15323.

Research of Texas A\&M has been supported by NSF grant AST-1814799.

GvdV acknowledges funding from the European Research Council (ERC) under the European Union's Horizon 2020 research and innovation programme under grant agreement No 724857 (Consolidator Grant ArcheoDyn).

\section*{Data Availability}

The HST data used in this work is publicly available at https://archive.stsci.edu/hst/ with project code 15323. The GEMINI data is available online via the GEMINI archive at https://archive.gemini.edu/searchform/, with project code GN-2019A-LP-8. Finally, the SAURON data is available via the Isaac Newton Group archive at http://casu.ast.cam.ac.uk/casuadc/ingarch/query identified with observation run numbers 958460, 958461, 958462, 958463, 958464, 958465 and 958466. The maps produced from these data can be shared on reasonable request to the corresponding author.

%%%%%%%%%%%%%%%%%%%%%%%%%%%%%%%%%%%%%%%%%%%%%%%%%%



\bibliographystyle{mnras}
\bibliography{references}

%%%%%%%%%%%%%%%%%%%%%%%%%%%%%%%%%%%%%%%%%%%%%%%%%%

%%%%%%%%%%%%%%%%% APPENDICES %%%%%%%%%%%%%%%%%%%%%

%%%%%%%%%%%%%%%%%%%%%%%%%%%%%%%%%%%%%%%%%%%%%%%%%%

\appendix

\section{Extinction and gas mass distribution in the polar ring}

\begin{figure}
    \centering
    \includegraphics[width=\columnwidth]{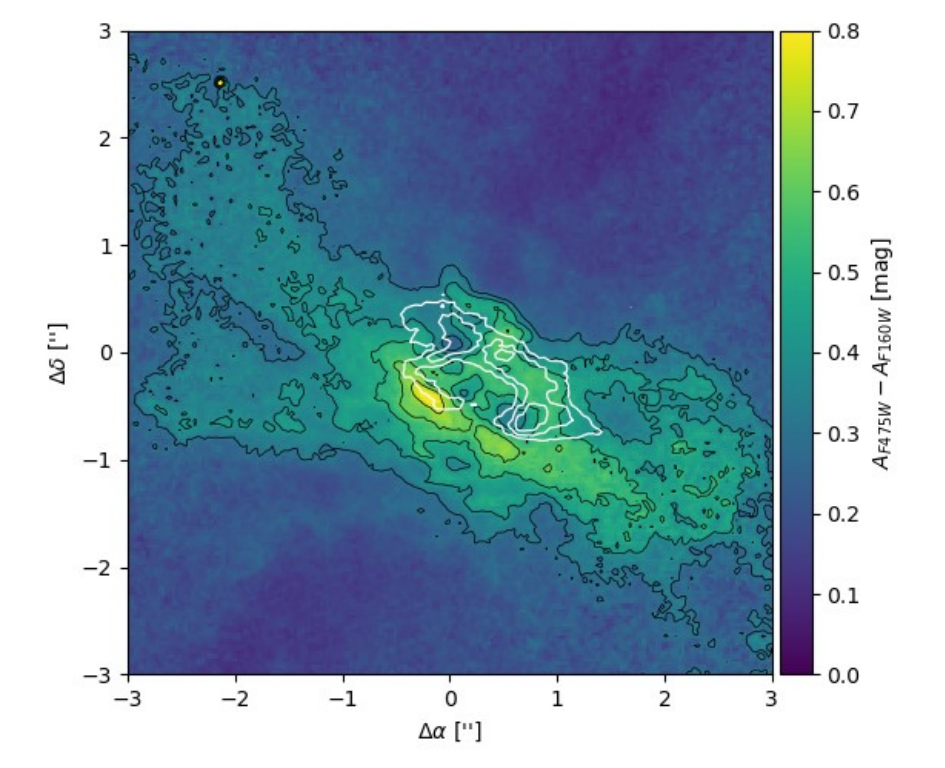}
    \caption{HST A$_{F475W}$ - A$_{F160W}$ map of the galaxy within the inner 6$^{\prime\prime}\times6^{\prime\prime}$ (440$\times$440\,pc$^2$), that covers the dusty polar ring. Overplotted in white are the contours from the H$_2 \lambda2.1218\,\mu$m emission line flux distribution obtained from the NIFS data.}
    \label{fig:av_map}
\end{figure}

From the colour map shown in Fig.\,\ref{fig:colourmap}, we have built the extinction map A$_{F475W}$ - A$_{F160W}$ shown in Fig. \ref{fig:av_map}, for the inner 6\arcsec$\times$6\arcsec\,(440$\times$440\,pc$^2$) of NGC\,4111, showing the reddening distribution along the dusty polar ring, calculated as described in Sec.\,\ref{ss:hst_images}. The spatial correlation with the H$_2\,\lambda2.1218\,\mu$m ring seen in the NIFS data is revealed by the overplotted white contours from the H$_2$ emission.

\begin{figure}
    \centering
    \includegraphics[width=\columnwidth]{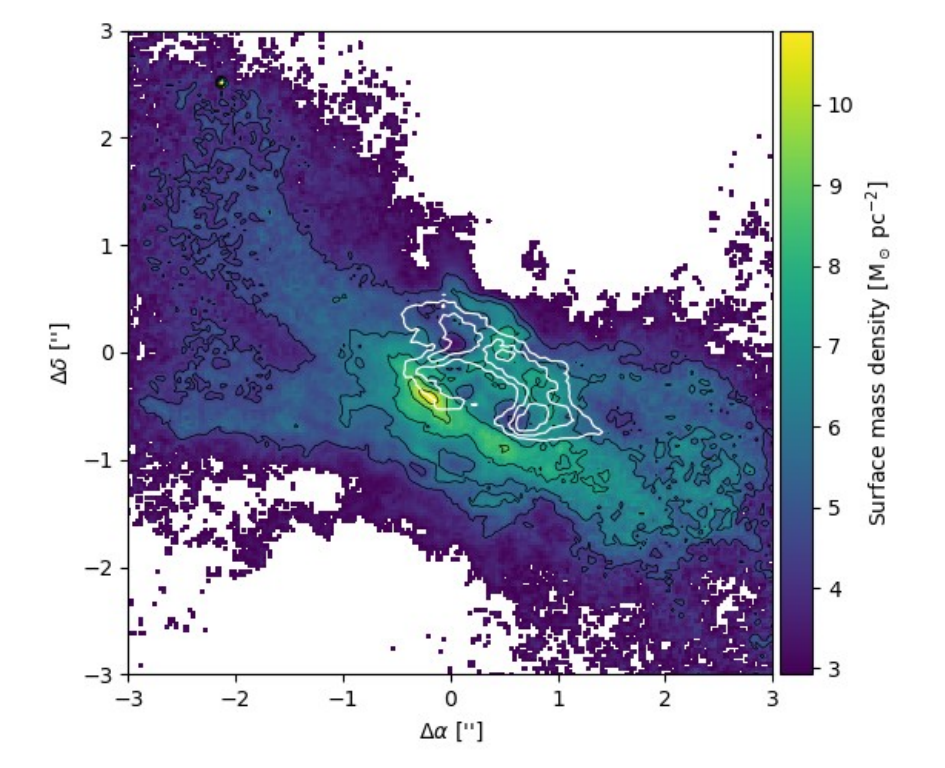}
    \caption{Gas mass column density map from the polar ring region obtained from the HST F475W - F160W colour map. Overplotted in white are the contours from the H$_2 \lambda2.1218\,\mu$m emission line flux distribution as derived from the NIFS data.}
    \label{fig:mass_hst}
\end{figure}

The second figure shown in this Appendix -- Fig. \ref{fig:mass_hst}, is a gas mass column density map of the same 6\arcsec$\times$6\arcsec\, (440$\times$440\,pc$^2$) region that covers the dusty polar ring. We again show in white, contours of the H$_2\,\lambda2.1218\,\mu$m line emission from NIFS data. This map has been obtained from the reddening map above following the steps discussed in Section \ref{ss:hst_images}. From this mass density map, integrating over the surface area of each spaxel, we obtained the cold gas mass as M$_{\text{gas}} = 9.8 \times 10^{6}$ M$_\odot$.

\section{Error maps from the NIFS kinematics}

\begin{figure*}
	\includegraphics[width=\textwidth]{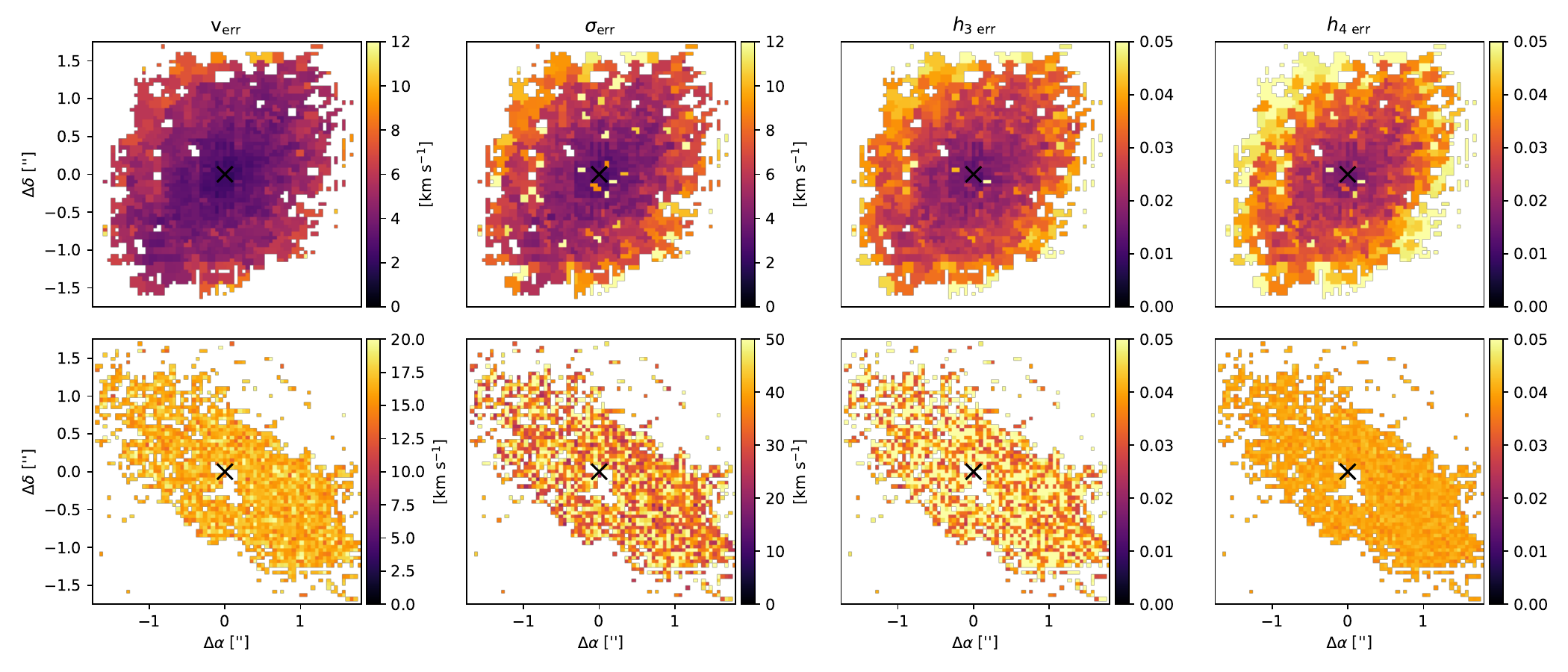}
    \caption{Monte Carlo error maps of the NIFS stellar (top panels) and molecular gas (bottom panels) kinematics fit. From left to right, the error maps of the velocity $v_\mathrm{err}$, velocity dispersion $\sigma_\mathrm{err}$ and Gauss-Hermite moments $h_{3\ \mathrm{err}}$ and $h_{4\ \mathrm{err}}$. The cross indicates the location of the galaxy nucleus.}
    \label{fig:nifs_kin_err}
\end{figure*}

In order to measure the intrinsic errors in the methods used for obtaining the NIFS stellar and molecular gas kinematics, we performed the Monte Carlo technique, perturbing the data with a Gaussian noise in every spaxel. From this, we obtained the error maps for the stellar and molecular gas kinematics, shown in Figure\,\ref{fig:nifs_kin_err}. Similarly to Figure\,\ref{fig:nifs_kin}, the top row shows the stellar kinematics errors and the bottom panels show those for the H$_2$ kinematics. From left to right, the panels show the error maps of the velocity $v_\mathrm{err}$, velocity dispersion $\sigma_\mathrm{err}$ and Gauss-Hermite moments $h_{3\ \mathrm{err}}$ and $h_{4\ \mathrm{err}}$. From these maps, we observe that the errors in the stellar kinematics decrease from the center to the borders, while those for the H$_2$ gas kinematics do not vary much. The mean errors of each kinematic parameter (of the measurements shown in each map, not considering the masked spaxels) are listed in Table\,\ref{tab:nifs_kin_err}.

\begin{table}
\caption{Mean Monte Carlo errors for each parameter of the NIFS stellar and molecular gas kinematics.}
\label{tab:nifs_kin_err}
\begin{center}
\begin{tabular}{llll}
\hline
\hline
Parameter             &   & Mean error  &     \\ \hline
%\multicolumn{3}{c}{Stellar:}{}             \\ 
        & &Stars          & Gas \\ \hline \hline
$v_\mathrm{err}$      &   & 4.9 km s$^{-1}$ & 17.9 km s$^{-1}$ \\
$\sigma_\mathrm{err}$ &   & 6.8 km s$^{-1}$ & 39.5 km s$^{-1}$ \\
$h_{3\ \mathrm{err}}$ &   & 0.03  & 0.05            \\
$h_{4\ \mathrm{err}}$ &   & 0.04  & 0.04            \\ \hline \hline
%\multicolumn{3}{c}{Molecular gas:}       \\ \hline
%$v_\mathrm{err}$      &   & 17.9 km s$^{-1}$ \\
%$\sigma_\mathrm{err}$ &   & 39.5 km s$^{-1}$ \\
%$h_{3\ \mathrm{err}}$ &   & 0.05             \\
%$h_{4\ \mathrm{err}}$ &   & 0.04             \\ \hline \hline
\end{tabular}
\end{center}
\end{table}

% Don't change these lines
\bsp	% typesetting comment
\label{lastpage}
\end{document}